\documentclass[seceqn]{elsart}

\usepackage{amsmath,latexsym,amssymb,tables,axodraw,psfig,bbm,graphicx,eufrak,mathtools}
\usepackage{macros,slashed,wick,wrapfig}
\usepackage[dvips]{epsfig}

\input eqalign.sty

\begin{document}

\begin{frontmatter}

\title{The nucleon electric dipole moment with the gradient flow: the $\theta$-term contribution}
\author[FZJ]{Andrea Shindler}
\author[FZJ]{Thomas Luu}
\author[FZJ]{Jordy de Vries}
\address[FZJ]{Institute~for~Advanced~Simulation, Institut~f\"{u}r~Kernphysik, 
J\"{u}lich~Center~for~Hadron~Physics,
JARA~HPC,\\
Forschungszentrum~J\"{u}lich,
D-52425~J\"{u}lich, Germany
}

\maketitle
\begin{abstract}
We propose a new method to calculate electric dipole moments induced by the strong QCD $\theta$-term.
The method is based on the gradient flow for gauge fields and is free from renormalization ambiguities.
We test our method by computing the nucleon electric dipole moments in pure Yang-Mills theory at several lattice spacings, 
enabling a first-of-its-kind continuum extrapolation. 
The method is rather general and can be applied for any quantity 
computed in a $\theta$ vacuum. 
This first application of the gradient flow has been successful and demonstrates proof-of-principle, 
thereby providing a novel method to obtain precise results for nucleon and light nuclear electric dipole moments. 
\end{abstract}

\end{frontmatter}
\cleardoublepage

\section{Introduction}
\label{seq:intro}

The electric dipole moments (EDMs) of the neutron and proton are very sensitive probes of CP-violating sources 
beyond those contained in the Standard Model (SM).
In fact, the current bound on the neutron EDM strongly constrains many models of  beyond-the-SM (BSM) physics. 
At current experimental accuracies, a nonzero nucleon EDM cannot be accounted for by the phase 
in the quark-mass matrix. 
This implies that such a signal is either caused by a nonzero QCD $\theta$ term or 
by genuine BSM physics which, at low energies, 
can be parametrized in terms of higher-dimensional CP-violating quark-gluon operators. 
Irrespective of the origin, the signal for the nucleon EDM will be small and largely 
masked by strong-interaction physics, which presents a formidable challenge to the interpretation
of such a signal.
To disentangle the origin of a nonzero EDM measurement (e.g. $\theta$-term or BSM), 
a quantitative understanding of the underlying hadronic physics is required.

The current experimental limit on the neutron EDM is $|d_N|<2.9 \cdot 10^{-13}~e \,{\rm fm}$~\cite{Baker:2006ts} 
and experiments are underway to improve this bound by one to two orders of magnitude.
The bound on the proton EDM is induced from the $^{199}{\rm Hg}$ EDM limit~\cite{Griffith:2009zz} 
and is $|d_P|<7.9 \cdot 10^{-12}~e\,{\rm fm}$. 
Plans exist to probe the EDM of the proton directly (and other light nuclei) in storage rings \cite{Pretz:2013us} with a 
proposed sensitivity of $10^{-16}~e \cdot {\rm fm}$, thus improving the current bounds 
by several orders of magnitudes and covering a wide range where BSM physics can show its footprint.

Nucleon EDMs arising from the QCD $\theta$-term or BSM physics have been calculated both in 
models~\cite{Pospelov:2005pr} 
and in chiral perturbation theory~\cite{Ottnad:2009jw,Mereghetti:2010kp}. 
In the latter approach, the nucleon EDMs are calculated 
in terms of effective CP-odd hadronic interactions that have the same symmetry properties 
as the underlying CP-odd sources at the quark level (for a review, see \cite{Mereghetti:2015rra}). 
The calculated EDMs depend on several low-energy constants (LECs) whose sizes are in most 
cases unknown and need to be estimated or calculated with lattice QCD.
 
Lattice QCD can thus be used to perform an {\it ab initio} calculation of the nucleon EDM.
For the $\theta$-term, this has already been shown in the pioneering works 
in refs.~\cite{Shintani:2005xg,Berruto:2005hg} and later in ref.~\cite{Shintani:2008nt}
(for BSM sources only the nucleon EDMs arising from the quark EDMs have been calculated 
with lattice QCD \cite{Bhattacharya:2015esa}).
The chiral and infinite volume extrapolations of unpublished lattice data from Shintani et al. 
have been performed in refs.~\cite{Guo:2012vf,Akan:2014yha}.
The calculation of the EDM within a lattice (discretized) formulation of QCD 
is very non-trivial, and present large difficulties for two main reasons. 
The renormalization of the CP-odd operators
and the degradation of the signal-to-noise ratio towards the chiral limit.
Additionally, the $\theta$-term itself introduces an imaginary term in the real Euclidean action,
which produces a sign problem and precludes the use of standard stochastic methods employed by lattice QCD.  
Ref. \cite{Guo:2015tla} performed a lattice QCD calculation of the neutron EDM 
induced by a $\theta$-term that was analytically continued into the complex plane.
This allows the usage of standard stochastic methods.

In this paper we propose, without relying on any complex rotation of the $\theta$-term,
a method based on the gradient flow for the gauge
fields~\cite{Luscher:2010iy} that has no renormalization ambiguities 
and, to our knowledge, is the only method that allows a theoretical sound continuum limit.
A first account of this method can be found in ref.~\cite{Shindler:2014oha}.

The remainder of the paper is organized as follows:  
The next section gives a cursory discussion of the phenomenology of the nucleon EDM. 
In sec.~\ref{sec:edm} we introduce definitions and our method.
Sec.~\ref{sec:gf} discusses the gradient flow for gauge fields and its relevance 
to the calculations presented in this paper.  
We provide details of our lattice calculations and their results in secs.~\ref{sec:2p} and \ref{sec:nedm}, 
followed by a discussion in the ensuing section.
 
\section{Phenomenology of the QCD theta term.}
\label{sec:phenom}

The discrete space-time symmetries parity P, time-reversal T, and the combination of charge conjugation
and parity CP, are broken in QCD by the QCD $\theta$ term. 
In the case of three quark flavors the QCD action is given by
\be
S_\theta = \int d^4 x~\left[\mcL_{\rm QCD}-i\theta q(x) \right]\,,
\label{eq:Stheta}
\ee
where $\mcL_{\rm QCD}$ is the standard Euclidean QCD Lagrangian 
\be
\mcL_{\rm QCD} = \frac{1}{4g^2}F_{\mu\nu}^aF^{a,\mu\nu} + \psibar(\gamma_\mu D_\mu + M)\psi
\label{eq:EQCD}
\ee
and
\be
q(x) = \frac{1}{64\pi^2} \epsilon_{\mu\nu\rho\sigma} F_{\mu\nu}^a(x)F_{\rho\sigma}^a(x)\,,
\label{eq:topocharge}
\ee
is the topological charge density.
The fermion field containing up, down and strange quarks is denoted by $\psi = (u\, ,  d\,, s)^T$ and
$F^a_{\mu\nu}$ is the gluon field strength tensor. 
$\epsilon_{\mu\nu\alpha\beta}$ ($\epsilon_{0123}=+1$)
is the completely antisymmetric tensor, $D_\mu$ the gauge-covariant derivative, $M$ the 
real $3\times 3$ quark-mass matrix, and $\theta$ the coupling
of the CP-odd interaction. 
In eq.~\eqref{eq:EQCD}  the complex phase of the quark-mass matrix has been absorbed 
in the physical parameter $\theta$, i.e. we choose
a fermionic basis where the CP-odd interaction comes solely from the topological charge density.

The most important consequence of the QCD $\theta$-term is that it induces EDMs of hadrons and nuclei. 
The first dedicated EDM experiment was the neutron EDM experiment in 1957 \cite{Purcell:1950zz}. 
Since then, the accuracy of the measurement has been improved by six orders of magnitude 
without finding a signal. 
The current bound $d_N<2.9\cdot 10^{-13} e \cdot$\, fm \cite{Baker:2006ts} sets strong 
limits on the size of $\theta$ and sources of $\CP$ violation from physics beyond the SM.

In order to set a bound on the $\theta$ term, it is necessary to calculate 
the dependence of the neutron EDM on $\theta$. 
One way to do this is by using $\chi$PT. 
This calculation has been done up to next-to-leading order (NLO) in both 
$SU(2)$ \cite{Hockings:2005cn,O'Connell:2005un,Mereghetti:2010kp}
and $SU(3)$ \cite{Crewther:1979pi,Borasoy:2000pq,Ottnad:2009jw} $\chi$PT. Focusing here on the two-flavored theory, 
the neutron ($d_N$) and proton EDM ($d_P$) are given by:
\be
d_N =  \frac{e g_A\bar{g}_0^\theta}{16\pi^2 F_\pi^2} \left(  \ln
\frac{M_\pi^2}{\Lambda_{N,\rm{EDM}}^2} -\frac{\pi M_\pi}{2 M_N} \right)\,,
\label{eq:pEDM}
\ee
\be
d_P = -\frac{e g_A \bar{g}_0^\theta}{16\pi^2 F_\pi^2}  \left(  \ln
\frac{M_\pi^2}{\Lambda_{P,\rm{EDM}}^2} -\frac{2\pi M_\pi}{M_N} \right) \,,
\label{eq:nEDM}
\ee
in terms of $g_A \simeq 1.27$ the strong pion-nucleon coupling constant, $F_\pi \simeq 92.2$ 
MeV the pion decay constant, $M_\pi$ and $M_N$ the pion and nucleon mass respectively, 
$e>0$ the proton charge, and, in principle, three low-energy constants (LECs) of 
CP-odd chiral interactions: $\bar{g}_0^\theta$ and $\bar d_{N,P}$. 
The first one, $\bar g_0^\theta$, is not free as discussed below (see eq.~\eqref{eq:g0theta}).
The latter two are absorbed in renormalization-scale, $\mu$, independent constants
\be
\Lambda_{N,\rm{EDM}} = \mu \exp\left\{-\frac{8 \pi^2 F_\pi^2 \bar{d}_N(\mu)}{e g_A \bar{g}_0^\theta}\right\}\,,
\ee
\be
\Lambda_{P,\rm{EDM}} = \mu \exp\left\{\frac{8 \pi^2 F_\pi^2 \bar{d}_P(\mu)}{e g_A \bar{g}_0^\theta}\right\}\,.
\ee
The first term in brackets in eqs.~\eqref{eq:pEDM} and \eqref{eq:nEDM} arises from the leading-order 
one-loop diagram involving the $\CP$-odd vertex 
\be
\mathcal L_{\pi N}(\theta) = -\frac{\bar{g}_0^\theta}{2 F_\pi}\, \bar N \vec \pi \cdot \vec \tau N\,,
\ee
in terms of the nucleon doublet $N$ and the pion triplet $\vec \pi$. 
The LO loop is divergent and the divergence and associated scale dependence 
have been absorbed into the counter terms $\bar d_{N,P}$ which signify contributions to the 
nucleon EDMs from short-range dynamics and appear at the same order as the LO loop diagrams. 
The second term in brackets in eqs.~(\ref{eq:pEDM}) and (\ref{eq:nEDM}) is a next-to-leading-order correction.

The $\theta$-term breaks chiral symmetry as a complex quark mass. As such, chiral symmetry relates $\bar g_0^\theta$ to
known CP-even LECs \cite{Crewther:1979pi,Mereghetti:2010tp}. In particular, it is possible to write \cite{deVries:2015una}
\be
\frac{\bar{g}^\theta_0}{2 F_\pi}= \left.\frac{(M_N -M_P)}{2 F_\pi}\right.^{\mathrm{strong}} \frac{m_\star}{\bar m \varepsilon} \theta = (15.5 \pm 2.5)\cdot 10^{-3}\,\theta\,\,\,,
\label{eq:g0theta}
\ee
where $(M_N - M_P)^{\mathrm{strong}}$ is the quark-mass induced part of the 
proton-neutron mass splitting, $\bar m = (m_u+m_d)/2$, $m_\star = m_u m_d/(m_u+m_d)$, and $\varepsilon = (m_u-m_d)/(m_u + m_d)$. 
To get a rough estimate of the sizes of the nucleon EDMs, we can insert
eq.~\eqref{eq:g0theta} in eqs.~(\ref{eq:nEDM}) and (\ref{eq:pEDM}). 
If we assume that $\Lambda_{\rm EDM}\simeq M_N$, we obtain
\be
d_N \simeq  -2.1\cdot 10^{-3}\,\theta\,e\,\mathrm{fm}\,,
\label{eq:nEDMnum}
\ee
\be
d_P \simeq   +2.5\cdot 10^{-3}\,\theta\,e\,\mathrm{fm}\,,
\label{eq:pEDMnum}
\ee
as a rough estimate of the nucleon EDMs. A comparison with the experimental bound then gives the strong constraint  $\theta \leq 10^{-10}$.  
Clearly, a more reliable constraint on $\theta$ requires a first-principle calculation of the nucleon EDMs. 
In the isospin limit,  $\bar{g}_0^{\theta}$  scales as $\bar m \sim M_\pi^2$ such that the loop contributions to the EDMs 
vanish in the chiral limit as $M_\pi^2 \log M_\pi^2$ (see eqs.~\ref{eq:pEDM} and \ref{eq:nEDM}).

In the isoscalar combination $d_N+d_P$ the loop contribution 
cancels out to a large extent. For observables sensitive to this combination, 
such as the deuteron EDM \cite{deVries:2011an,Bsaisou:2012rg}, a first-principle calculation 
of the total nucleon EDM is important to differentiate the $\theta$-term from possible 
BSM sources of CP violation \cite{Lebedev:2004va,Dekens:2014jka}. 
In the specific case of the isoscalar combination
a precise evaluation of disconnected diagrams is needed in any lattice QCD calculation.

\section{The electric dipole moment}
\label{sec:edm}

The theory is defined in Euclidean space with the action given in eq.~\eqref{eq:Stheta}.
The EDM of a nucleon is related to the spatial charge density distribution.
If we define the quark charges as $Q_u=2/3~e$ and $Q_d=Q_s=-1/3~e$, 
the nucleon EDMs are obtained from the matrix element of the electromagnetic current 
\be
J_\mu(x) = Q_u \ubar(x)\gamma_\mu u(x) + Q_d \dbar(x)\gamma_\mu d(x) + Q_s \sbar(x)\gamma_\mu s(x)\,,
\ee
between nucleon states in the $\theta$ vacuum,
\be
\langle N^\theta(\bp',s')| J_\mu | N^\theta(\bp,s) \rangle = 
\bar{u}_N^\theta(\bp',s') \Gamma_\mu(Q^2) u_N^\theta(\bp,s)\,.
\label{eq:ampli}
\ee
$\Gamma_\mu(Q^2)$ has the most general four-vector
structure consistent with the symmetries of the action~\eqref{eq:Stheta} such as 
gauge, O(4), C and CPT invariance .
Note that the photon momentum $q=p'-p$ in Euclidean space is 
\be
Q_\mu= (Q_4,\bQ) = (iq^0,\bq)\,, Q^2 = -(q^0)^2 + |\bq|^2 = - q^2\,.
\ee
Following ref.~\cite{Shintani:2005xg} 
the $Q^2$ dependence of the matrix element is parametrized by a linear combination of 
CP-even and CP-odd form factors. Using Euclidean O($4$) rotational invariance, gauge symmetry and 
the spurionic symmetry $P \times \theta \rightarrow -\theta$,
the most general decomposition reads
\be
\Gamma_\mu(Q^2) = g(\theta^2)\Gamma^{\rm even}_\mu(Q^2)+i\theta h(\theta^2) \Gamma_\mu^{\rm odd}(Q^2)\,,
\label{eq:gamma_Q2}
\ee
where $g,h$ are even functions of $\theta$. The CP-even contribution is given by
\be
\Gamma^{\rm even}_\mu(Q^2) = \gamma_\mu F_1(Q^2) + \sigma_{\mu\nu} \frac{Q_\nu}{2M}F_2(Q^2) 
\label{eq:gamma_evenQ2}
\ee
where the Dirac and Pauli form factors $F_1$ and $F_2$ are related to the electric
and the magnetic form factors
\be
G_E(Q^2) = F_1(Q^2) -\frac{Q^2}{4M^2}F_2(Q^2)\,, \qquad 
G_M(Q^2)=F_1(Q^2)+F_2(Q^2)\,.
\ee
The CP-odd term reads
\be
\Gamma^{\rm odd}_\mu(Q^2)= \sigma_{\mu\nu}\gamma_5 \frac{Q_\nu}{2M}F_3(Q^2)\,.
\label{eq:gamma_oddQ2}
\ee
In the literature $\Gamma_\mu^{\rm odd}(Q^2)$ usually contains an additional parity violating form factor, 
the anapole form factor.
The anapole form factor breaks parity symmetry but does not break time reversal, i.e. is CP-even
while breaking both C and P.
It therefore does not contribute to the amplitude in eq.~\eqref{eq:ampli} of the electromagnetic current
evaluated in a $\theta$ vacuum. In other words, the $\theta$-term alone cannot induce a nucleon anapole form factor.
The EDM is directly related to the CP-odd $F_3(Q^2)$ form factor at zero momentum transfer
\be
d_N = \theta g(\theta^2)\frac{F_3^N(0)}{2M_N}\simeq \theta \frac{F_3^N(0)}{2M_N}+O(\theta^3)\,.
\ee

In lattice calculations, matrix elements can be extracted from the
large-distance behavior of appropriate correlation functions in Euclidean
space-time. In the case at hand, one considers three-point correlations such as
\be
G^{\theta}_{NJ_\mu N} = \langle \mcN J_\mu \bar{\mcN} \rangle_\theta\,,
\ee
where $\mcN$ is an interpolating operator with the same quantum number of the nucleon.
The three-point functions are to be evaluated with the Euclidean action $S_\theta$. 
Although the action with $\theta\ne 0$ cannot be directly studied by numerical
Monte Carlo methods, in the small $\theta$ limit one can obtain the desired
result for EDM by expanding around $\theta=0$ and taking
only the linear term in $\theta$. That is, for a generic expectation value of product of
operators, $\mathcal{O}$, in a $\theta$-vacuum, we can write
\be
\langle  \mcO \rangle_\theta \simeq 
\langle  \mcO \rangle_{\theta=0} + i \theta \langle \mcO \int d^4x~ q(x) \rangle_{\theta=0} + {\rm O}(\theta^2)\,,
\label{eq:O_theta}
\ee
where $q(x)$ is the topological charge density~\eqref{eq:topocharge}.
The experimental bound on $\theta$ is currently $\theta < O(10^{10})$ (see sec.~\ref{sec:phenom}),
thus a power expansion in $\theta$ is well justified\footnote{
Alternatively, the nucleon EDM at finite $\theta$ can be 
also determined using reweighting techniques with
the complex weight factor $e^{i\theta Q}$.}.

In general this proposal could be hampered
by the impossibility of giving a sound or practical definition on the lattice of the topological charge
density and its continuum limit.
In this work we propose to directly compute the matrix element
\be
\langle \mcO \int d^4x~ q(x) \rangle_{\theta=0}\,,
\ee
using the gradient flow (see sect.~\ref{sec:gf}) to define the topological charge density~\cite{Luscher:2010iy}. 
By doing so, we have a theoretically sound definition of the correlation function with 
no renormalization ambiguities and a well-defined continuum limit.

\section{Gradient flow and the topological susceptibility}
\label{sec:gf}

The gradient flow~\cite{Luscher:2010iy} of Yang-Mills gauge fields is defined as follows
\be
\partial_t B_{\mu}=D_{\nu,t}G_{\nu\mu}\,,
\label{eq:GF_gauge}
\ee
where the flow-time $t$ has a time-squared dimension, 
\be
G_{\mu\nu}=\partial_{\mu}B_{\nu}-\partial_{\nu}B_{\mu}+[B_{\mu},B_{\nu}]\,,
\qquad
D_{\mu,t}=\partial_{\mu}+[B_{\mu},\,\cdot\;]\,,
\label{eq:Gmunu_t}
\ee
and the initial condition on the flow-time-dependent field $B_{\mu}(t,x)$
at $t=0$ is  given by the fundamental gauge field.
\begin{figure}
\begin{center}
\includegraphics[width=10cm]{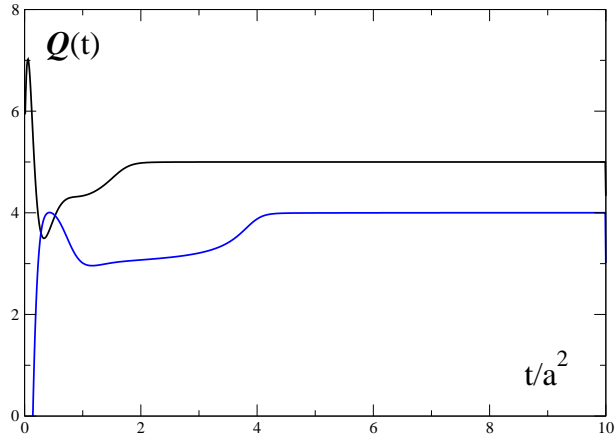}
\caption{Flow-time dependence of the topological charge for two different gauge fields.}
\label{fig:Q_t}
\end{center}
\end{figure}
The gradient flow for gauge fields and for fermions~\cite{Luscher:2013cpa}
has several applications and here we mention the definition of a relative scale~\cite{Luscher:2010iy,Borsanyi:2012zs},
the determination of the strong coupling constant~\cite{Luscher:2010iy,Fritzsch:2013je} 
and of the chiral condensate~\cite{Luscher:2013cpa,Shindler:2013bia}, 
the calculation of the energy-momentum tensor~\cite{Suzuki:2013gza,DelDebbio:2013zaa} 
and of the topological susceptibility~\cite{Chowdhury:2013mea,Bruno:2014ova}.
We have recently proposed to use the gradient flow for the determination 
of the strange content of the nucleon~\cite{Shindler:2014oha}.

One way to understand the flow equations is to consider them as steepest descent equations
in the space of gauge fields. As such the evolution along the flow drives the 
gauge configurations towards local minima of the action. The topological charge 
is defined at non-vanishing flow-time as
\be
\mcQ(t) = \int d^4 x~q(x,t)\,,
\label{eq:Qdef}
\ee
with
\be
q(x,t) = \frac{1}{64\pi^2} \epsilon_{\mu\nu\rho\sigma} G_{\mu\nu}^a(x,t)G_{\rho\sigma}^a(x,t)\,.
\label{eq:topo_density_t}
\ee
In fig.~\ref{fig:Q_t} we show the flow-time
evolution of $\mcQ(t)$ evaluated on two representatives of our gauge ensembles.
Rather rapidly $\mcQ(t)$ reaches a plateau where it assumes an almost integer value saturating the 
corresponding instanton bound. 

Another way to understand the effect of the gradient flow on the gauge fields is apparent already at tree-level.
The smoothing at short distances over a range $\sqrt{8t}$ corresponds in momentum space to a Gaussian damping
of the large momenta. This results in a very interesting property of the flowed gauge fields $B_\mu(x,t)$:
they are free from ultraviolet divergences~\cite{Luscher:2010iy,Luscher:2011bx} for all $t>0$ 
and do not require any renormalization.
This powerful result can be used to simplify the renormalization pattern
of operators involving gauge fields. 
In general, one would need to relate the local operators
evaluated at non-vanishing flow-times with the ones at zero flow-times.
The case of the topological charge and all the correlation functions
containing the topological charge is special, because in this case we can define the
topological charge, and for example the topological susceptibility, directly at 
non-vanishing flow-time~\cite{Luscher:2010iy}.
\begin{wraptable}{r}{6.8cm}
\begin{center}
\begin{tabular}{|c|c|c|c|c|} 
\hline 
$\beta$ & $N_{\rm th}$ & $N_{\rm up}$ & $N_g$ & $N_{\rm meas}$ \\
\hline
6.0 & 2000 & 200000 & 1000 & 1000 \\
6.1 & 2000 & 65000 & 325 & 325 \\
6.2 & 2000 & 60000 & 300 & 300 \\
6.45 & 2000 & 122400 & 612 & 153 \\
\hline
\end{tabular} 
\caption{Summary of our runs: $N_{\rm th}$ is the number of thermalization updates; $N_{\rm up}$ is the total number of updates; 
$N_g$ is the number of gauges saved and $N_{\rm meas}$ is the number of gauges analyzed.}
\label{tab:runs} 
\end{center}
\end{wraptable}
The Euclidean theory is prepared on a lattice of spacing $a$ and volume $L^3 \times T$. 
The calculations in this and the following sections have been performed with the standard Wilson gauge
action, with $\beta=6/g^2$, at $4$ different lattice spacings $a=0.093, 0.079, 0.068, 0.048$ fm corresponding
to $\beta=6.0, 6.1, 6.2, 6.45$. In this work we use the Sommer scale~\cite{Sommer:1993ce,Necco:2001xg}, 
$r_0=0.5$ fm, to fix the lattice spacing in physical units.
The size of the box is respectively $L/a=16, 24, 24, 32$ with $T/L=2$.
To generate the gauge links, we have used a heat bath algorithm
with a ratio of number of over-relaxation steps, $N_{OR}$, over a number of Cabibbo-Marinari updates, $N_{HB}$, per sweep
of $N_{OR}/N_{HB}=4/1$. 
For thermalization we have performed $2000$ updates. 
For the finest lattice spacing, we have analyzed all correlation functions skipping $800$ gauges
while for the remaining correlation functions we have skipped $200$ gauges.
A summary of parameter runs is given in tab.~\ref{tab:runs}.

With these choices we have observed no significant autocorrelation for all
our lattice spacings. We obtain the same outcome also for the 
correlation functions used for the determination of the EDMs. 
A more detailed discussion of autocorrelations for the fermionic correlation functions is given in 
sec.~\ref{sec:2p}.

The gradient flow equation at finite lattice spacing is solved following app. C 
of ref.~\cite{Luscher:2010iy} with step-size for the flow-time $\epsilon=0.01$.
The topological charge density is defined as in eq.~\eqref{eq:topo_density_t}
where $G_{\mu\nu}^a(x,t)$ is the lattice implementation of the field tensor defined in 
ref.~\cite{BilsonThompson:2002jk}. Any other definition of the topological
charge density in a pure Yang-Mills theory 
requires a finite multiplicative renormalization~\cite{Vicari:2008jw} that has to be determined 
as a function of the bare coupling, in order to perform the continuum limit.
With the definition based on the gradient flow, this renormalization factor is $1$
independently of the lattice action used.
\begin{figure}
\includegraphics[width=8cm]{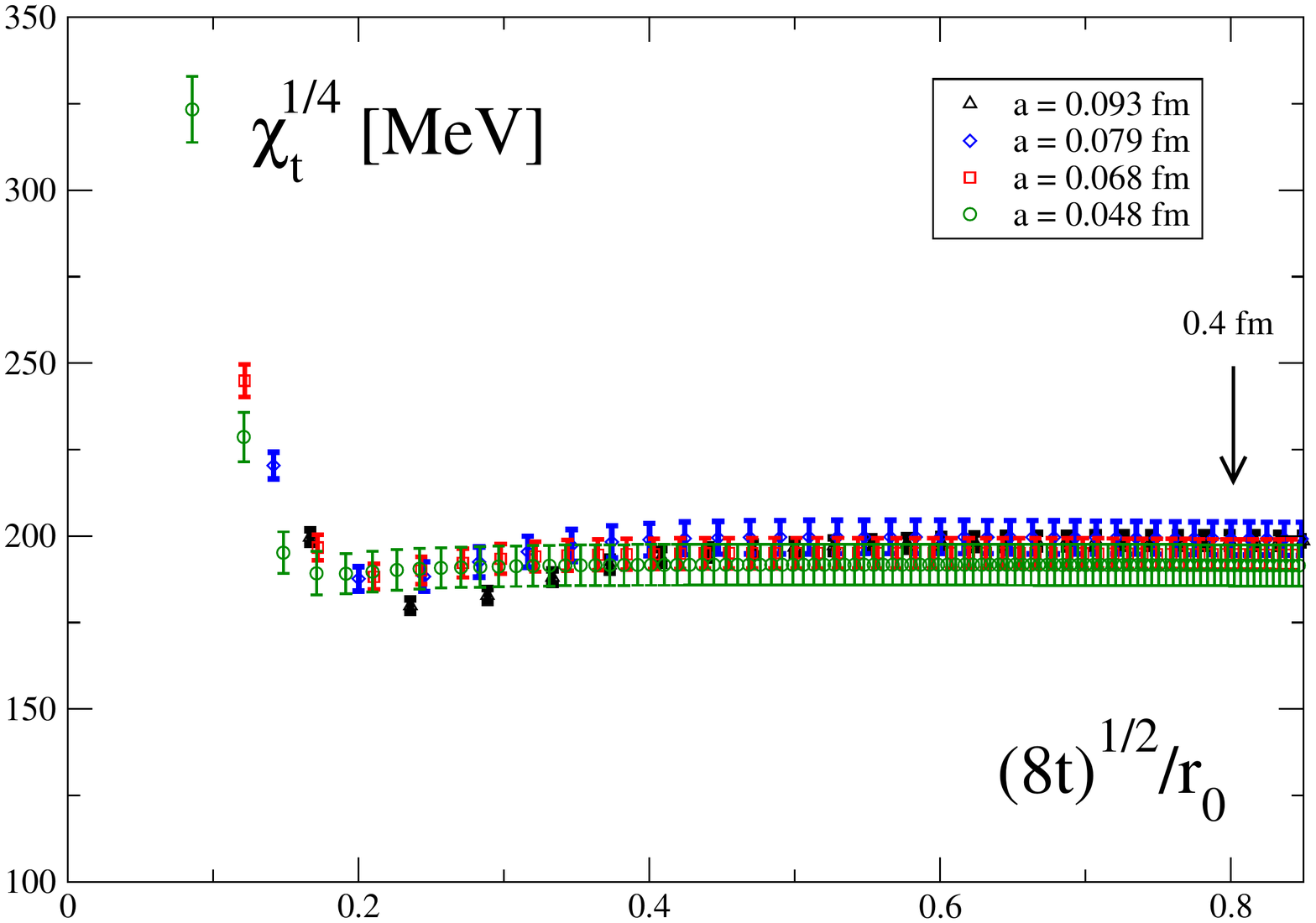}
\includegraphics[width=8cm]{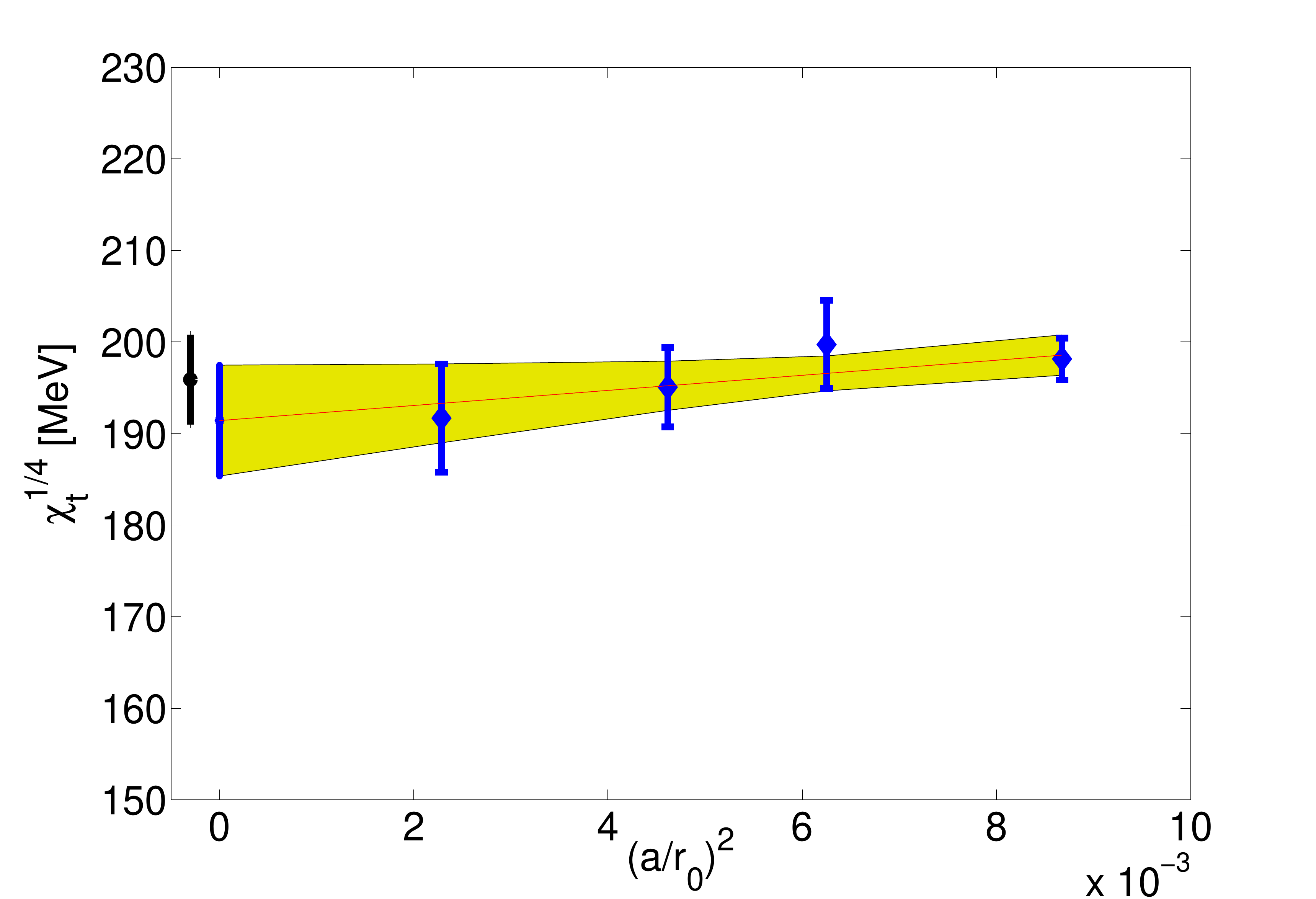}
\caption{Left plot: flow-time dependence of the topological susceptibility at several lattice spacings.
Right plot: continuum limit of the topological susceptibility. The yellow band is a linear extrapolation in $a^2$
compared with a constant fit.}
\label{fig:chi}
\end{figure}
Using the definition of the topological charge density at non-vanishing flow-time given above, 
we can also define the topological susceptibility 
\be
\chi_{\rm t}(t) = \frac{1}{V}\int d^4x~d^4y \left\langle q(t,x) q(t,y) \right\rangle \,.
\label{eq:topo_susc}
\ee
The topological susceptibility defined as in eq.~\eqref{eq:topo_susc}, but at vanishing flow-time $t=0$,
not only needs a multiplicatively renormalization, but, more importantly, has a $1/a^4$ power divergence
\footnote{A notable exception is the definition of the topological susceptibility proposed in ref.~\cite{Giusti:2008vb}
based on spectral projectors.}.
However, with the definition at non-vanishing flow-time, the topological susceptibility needs no renormalization
and it has a well defined continuum limit.
We have computed $\chi_t$ for several lattice spacings as a function of the flow-time. 
In the left plot of fig.~\ref{fig:chi} we show the topological susceptibility 
in physical units as a function of $\sqrt{8t}/r_0$. 
The divergence-free property of the gauge fields at non-vanishing flow time allows us 
to perform the continuum limit at fixed value of $\sqrt{8t}/r_0$.  
In the continuum limit we expect the topological susceptibility to be flow-time independent
for every positive flow-time, $\sqrt{8t}/r_0 > 0$~\cite{Luscher:2010iy}. 

For small flow-time values we observe two different effects. 
First, for $\sqrt{8t} \lesssim 0.1$ fm we observe a rapid increase of $\chi_t$  
that is just a reflection of the short distance singularities discussed above. 
Second, for $0.1$ fm $\lesssim \sqrt{8t} \lesssim 0.2$ fm we observe
some discretization effects. For $\sqrt{8t} > 0.2$ fm we find complete agreement 
between all lattice spacings and, as expected, $\chi_t$ is flow-time independent.
We perform the continuum limit at $\sqrt{8t}/r_0=0.8$ and this is shown in the right plot of
fig.~\ref{fig:chi} where we compare a linear extrapolation in $a^2$ with a constant one.
We decide to quote as final result
\be
\left[\chi_{\rm t}\right]^{1/4} = 195.9(4.9) {\rm~MeV}\,,
\ee
that is a constant fit including all lattice spacings.
The values at all lattice spacings and different extrapolations to the continuum limit are listed in tab.~\ref{tab:cl_chi_alpha}.
This result is in perfect agreement with the result~\cite{DelDebbio:2004ns} obtained using the index theorem 
with a chiral lattice Dirac operator and the result~\cite{Luscher:2010ik} obtained using the spectral projector method.
Very recently a paper has been submitted with a precise determination 
of the topological susceptibility~\cite{Ce:2015qha} using the gradient flow. 
The results are consistent within statistical uncertainties.

\section{CP-broken vacuum and nucleon mixing}
\label{sec:2p}

The form factor $F_3$, directly related to the nucleon EDM, defined in eq.~\eqref{eq:gamma_oddQ2} 
can be computed non-perturbatively with suitable ratios of the following 2- and 3-point functions in a $\theta$ vacuum
\be
G_{NN}^\theta(\bp,x_0) = a^3 \sum_{\bx} \e^{i\bp\bx} \left\langle  \mcN(\bx,x_0) \overline{\mcN}(0)\right\rangle_\theta\,, 
\ee
\be
G_{N J_\mu N}^\theta(\bp_1,\bp_2,x_0,y_0) = a^6 \sum_{\bx,\by} \e^{i\bp_2(\bx-\by)}\e^{i\bp_1\by} \left\langle \mcN(\bx,x_0)  J_\mu(\by,y_0)  \overline{\mcN}(0)\right\rangle_\theta\,.
\ee
Here, the baryon interpolating fields are 
\be
\mcN(x) = \varepsilon_{ABC} u_A(x)\left[ u^T_B(x) \mcC \gamma_5 d_C(x)\right]\,,
\ee
\be
\overline{\mcN}(x) = \varepsilon_{ABC} \left[ \ubar^T_A(x) \mcC \gamma_5 \dbar_B^T(x)\right]\ubar_C(x)\,,
\ee
and $\mcC$ is the charge conjugation matrix.
We now describe in some detail the spectral decomposition for the 2-point functions and defer to app.~\ref{app:A} 
for the slightly more cumbersome spectral decomposition of the 3-point functions. 
Most of the discussion on the spectral decomposition follows Shintani et al.~\cite{Shintani:2005xg}, but we 
rederive some of their results for clarity and to be consistent with our normalizations.
The key ingredient of the spectral decompositions is the matrix element of the interpolating operator 
of the nucleon between the $\theta$ vacuum and a single nucleon state
\be
\left\langle \theta | \mcN|N^\theta(\bp,s)\right\rangle\,=\mcZ_N(\theta) u_N^\theta(\bp,s)\,.
\ee
In a theory that does not preserve parity, for instance due to the presence of a $\theta$-term, 
the nucleon state does not have a definite parity and the nucleon spinor can be written 
as 
\be
u_N^\theta(\bp,s) = \e^{i\alpha_N(\theta)\gamma_5}u_N(\bp,s)\,,
\ee
where $u_N(\bp,s)$ is the nucleon spinor in the $\theta=0$ vacuum.
In other words, the nucleon spinor satisfies the modified Dirac equation 
\be
\left(i \gamma_\mu p_\mu + M_N(\theta) \e^{- i 2\alpha_N(\theta)\gamma_5}\right)u_N^\theta(\bp,s) =0\,.
\ee
The theory still preserves the spurionic symmetry $P_\theta:~P \times \theta \rightarrow -\theta$, where $P$ is the 
standard parity transformation. This implies that both the energies and the amplitudes $M(\theta), \mcZ(\theta)$
are even functions of $\theta$, $M(\theta)=M+O(\theta^2)$ and $\mcZ_N(\theta)=\mcZ_N+O(\theta^2)$.

The phase $\alpha_N(\theta)$ plays a very important role in the determination of the EDM.
From the spurionic symmetry $P_\theta$ we deduce that $\alpha_N(\theta) = -\alpha_N(-\theta)$
and for small values of $\theta$, $\alpha_N(\theta) = \alpha_N^{(1)}\theta + O(\theta^3)$.
It is important to determine precisely the mixing parameter $\alpha_N$ before extracting the CP-odd
form factors from the 3-point functions. The reason is that the mixing between different parity states 
can induce a spurious CP-odd contribution to the correlation function
proportional to the CP-even form factors.
These spurious contributions can be subtracted only with a precise
determination of the mixing angle $\alpha_N(\theta)$. 
The details of these spurious contributions and relative subtractions are detailed in app.~\ref{app:A}.

For on-shell nucleons with energy $-ip_0=E_N(\bp)$ where $E_N(\bp)=\sqrt{\left|\bp\right|^2 + M_N^2}$, 
the infinite volume normalization reads
\be
\langle N^\theta(\bq,s)|N^\theta(\bk,s')\rangle = (2\pi)^3\sqrt{2E_N(\theta;\bq)}\sqrt{2E_N(\theta;\bk)}\delta^{(3)}(\bk-\bq)\delta_{s,s'}\,.
\ee
Taking into account the parity mixing, the completeness relation of the nucleon spinors
with spatial momentum $\bp$ reads
\be
\sum_s u_N^\theta(\bp,s) \ubar_N^\theta(\bp,s) = 
E_N(\theta;\bp)\gamma_0 - i \gamma_k p_k +M_N(\theta) {\rm e}^{2 i \alpha_N(\theta)\gamma_5}\,.
\ee
For small values of $\theta$, we have 
\be
\sum_s u_N^\theta(\bp,s) \ubar_N^\theta(\bp,s) = 
E_N(\bp)\gamma_0 - i \gamma_k p_k +M_N \left(1+ 2 i \theta \alpha_N^{(1)}\theta\gamma_5\right) + O(\theta^2)\,.
\ee
We can now perform the spectral decomposition of the nucleon 2-point functions in a $\theta$-vacuum.
Retaining only the one-state leading contribution we obtain
\be
G_{NN}^\theta(\bp,x_0) = 
\frac{{\rm e}^{-E_N(\theta;\bp)x_0}}{2 E_N(\theta;\bp)} \left|\mcZ_N(\theta;\bp)\right|^2 
\sum_s u_N^\theta(\bp,s) \ubar_N^\theta(\bp,s)\,,
\ee
and using the completeness relation we get
\be
G_{NN}^\theta(\bp,x_0)_{\alpha\beta} = \frac{{\rm e}^{-E_N(\theta;\bp)x_0}}{2 E_N(\theta;\bp)} 
\left|\mcZ_N(\theta;\bp)\right|^2 
\left[E_N(\theta;\bp)\gamma_0 -i\gamma_k p_k + M_N(\theta)\e^{2 i \alpha_N(\theta)\gamma_5}\right]_{\alpha\beta}\,,
\label{eq:SD1}
\ee
where $\alpha$ and $\beta$ are the Dirac indices.
Expanding the l.h.s. of eq.~\eqref{eq:SD1} in powers of $\theta$, we obtain
\be
G_{NN}^\theta(\bp,x_0) = G_{NN}(\bp,x_0) + i \theta G_{NN}^\mcQ(\bp,x_0) + {\rm O}(\theta^2)
\ee
where
\be
G_{NN}(\bp,x_0) = a^3 \sum_{\bx} \e^{i\bp \bx}\left\langle \mcN(\bx,x_0) \overline{\mcN}(0) \right\rangle\,,
\label{eq:SDNN}
\ee
and
\be
G_{NN}^{\mcQ}(\bp,x_0) = a^3 \sum_{\bx} \e^{i\bp \bx} \left\langle \mcN(\bx,x_0)  \overline{\mcN}(0) \mcQ\right\rangle\,.
\label{eq:SDNNQ}
\ee
The term linear in $\theta$ can be computed inserting the topological charge in the nucleon 
2-point function. The topological charge $\mcQ$, defined in eq.~\eqref{eq:Qdef}, 
is computed as detailed in sect.~\ref{sec:gf} using the gradient flow.
In this way the topological charge is free from any renormalization ambiguity and the continuum limit
can be safely performed keeping fixed the flow-time in physical units.
To minimize discretization effects we choose, $\sqrt{8t}/r_0=0.8$. We omit the flow-time dependence
of $\mcQ(t)$ because in this range of flow-times any correlator involving the topological
charge is flow time independent (see sect~\ref{sec:gf}).

By expanding the spectral decomposition, i.e. the r.h.s. of eq.~\eqref{eq:SD1},  in powers of $\theta$,
we obtain the standard nucleon spectral decomposition
\be
G_{NN}(\bp,x_0) = \frac{{\rm e}^{-E_N(\bp)x_0}}{2 E_N(\bp)} 
\left|\mcZ_N(\bp)\right|^2 
\left[E_N(\bp)\gamma_0 -i\gamma_k p_k + M\right]\,,
\ee
and the term linear in $\theta$
\be
G_{NN}^{\mcQ}(\bp,x_0) = \frac{{\rm e}^{-E_N(\bp)x_0}}{2 E_N(\bp)} 
\left|\mcZ_N(\bp)\right|^2 
2 M_N \alpha_N^{(1)} \gamma_5 \,.
\ee
For simplicity we have not written down the opposite parity states propagating from $T$.
If we project to $\bp=\bzero$ and to positive parity we obtain
\be
C(x_0)={\rm tr}\left[P_+G_{NN}(\bzero,x_0)\right] = 2 |Z_N|^2 {\rm e}^{-M_N x_0} + \cdots\,,
\label{eq:GNN}
\ee
and
\be
C^\mcQ(x_0)={\rm tr}\left[P_+ \gamma_5 G_{NN}^\mcQ(\bzero,x_0)\right] =  2 |Z_N|^2 \alpha_N^{(1)} {\rm e}^{-M_N x_0} + \cdots\,.
\label{eq:GNNQ}
\ee
We observe that the two correlators have the same leading exponential behavior. If the sampling of all
topological sectors is correctly performed, the effective masses of the two correlators should 
agree asymptotically for large Euclidean times.
\begin{figure}
\begin{center}
\includegraphics[width=7cm]{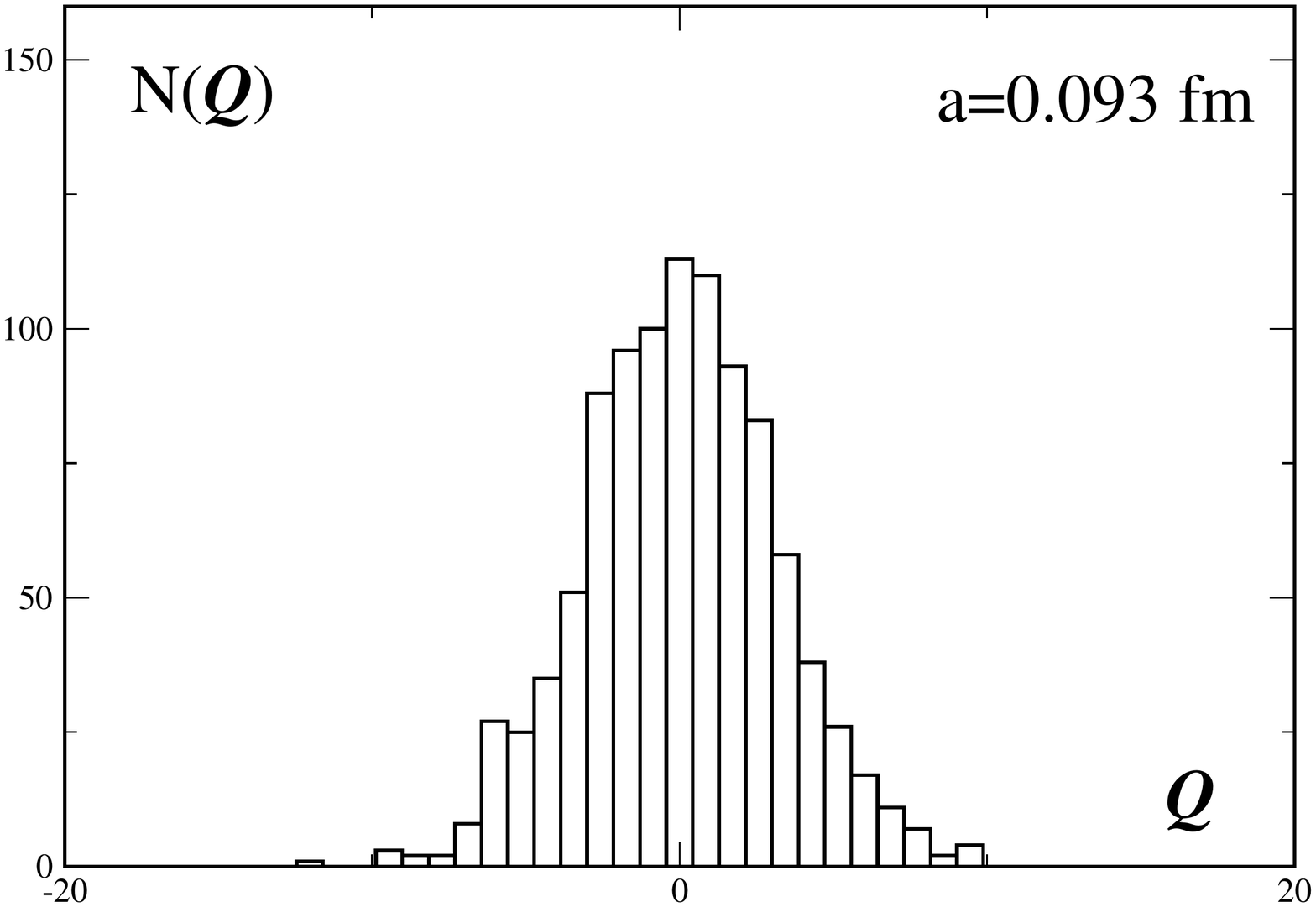}
\includegraphics[width=7cm]{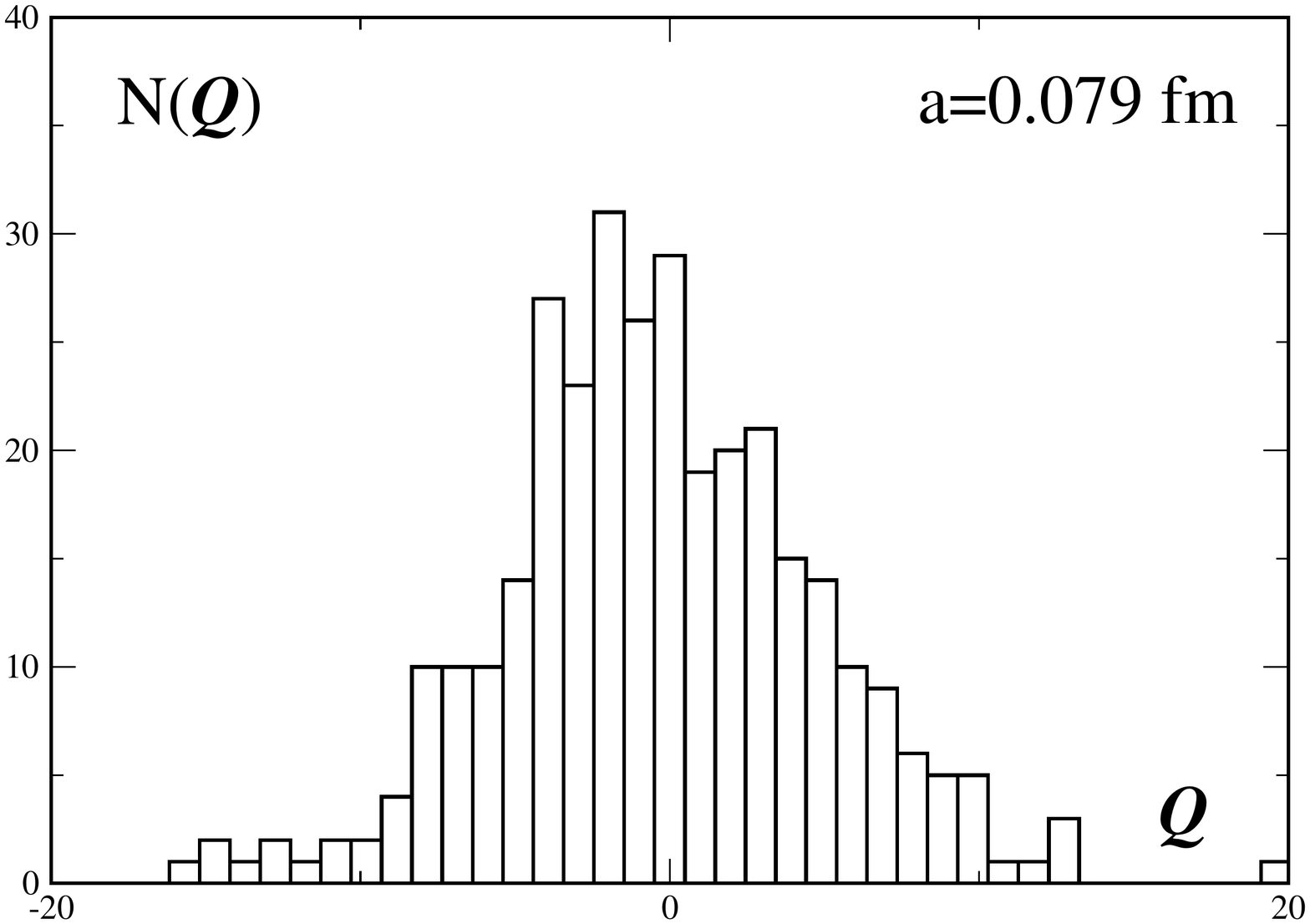} \\
\end{center}
\begin{center}
\includegraphics[width=7cm]{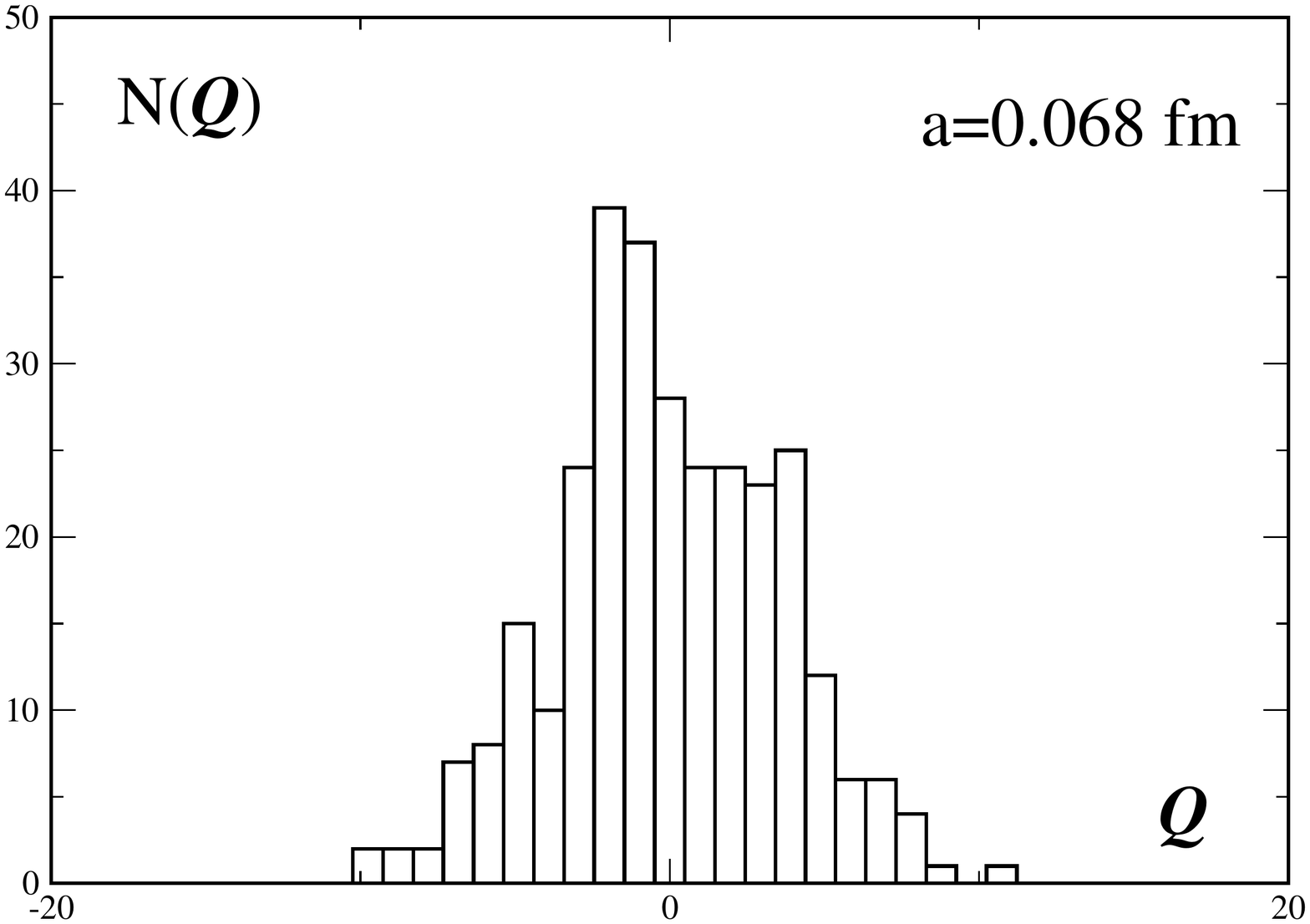}
\includegraphics[width=7cm]{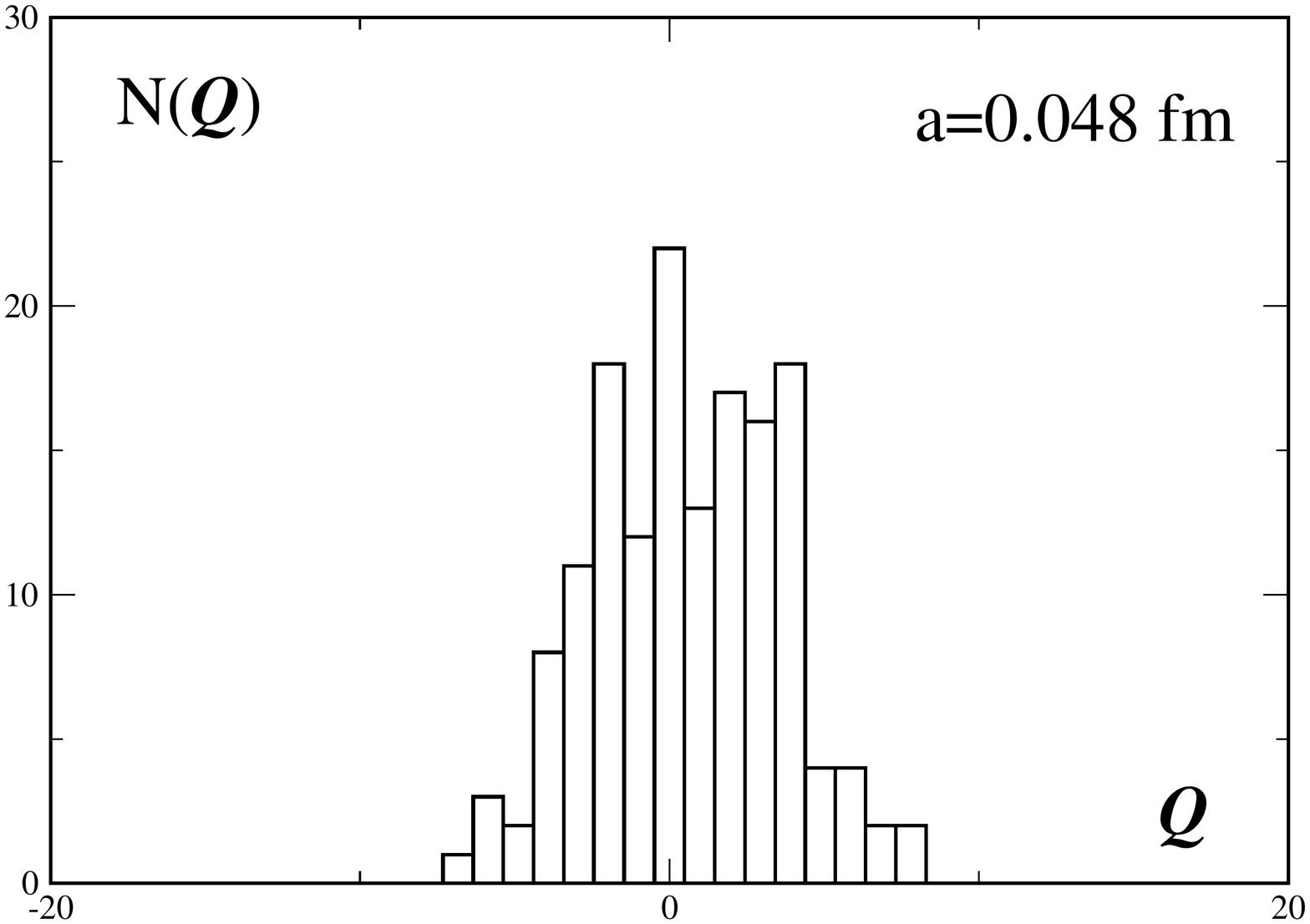}
\end{center}
\caption{Distribution of the topological charge computed for $\sqrt{8t}/r_0=0.8$ for 4 different lattice spacings.}
\label{fig:topo_distr}
\end{figure}

\begin{table}
\begin{center}
\begin{tabular}{|c|c|c|} 
\hline 
$a^2 \cdot 10^3~\left[{\rm fm}^2\right]$ & $\chi_t^{1/4}\left[{\rm MeV}\right]$ & $\alpha_N$ \\
\hline
8.675135 &  198.1(2.3) &  0.289(19) \\
6.250475 &  199.7(4.8) &  0.314(38) \\
4.615747 &  195.1(4.3) &  0.324(33) \\
2.285814 &  191.7(5.9) &  0.301(35) \\
 0~~ [{\rm fit~}1] &  195.9(4.9) & 0.314(35) \\
 0~~ [{\rm fit~}2] &  191.4(6.0) & 0.326(40) \\
\hline
\end{tabular} 
\caption{Numerical results for the topological susceptibility and the CP-odd mixing angle $\alpha_N^{(1)}$ 
for several lattice spacings. The continuum extrapolated values are obtained with a constant fit using the 
3 finest lattice spacings (fit 1) and a linear extrapolation in $a^2$ using all the lattice spacings (fit 2).}
\label{tab:cl_chi_alpha} 
\end{center}
\end{table}
In fig.~\ref{fig:topo_distr} we plot the distribution of the topological charge for 4 different lattice spacings
at $\sqrt{8t}=0.8 r_0$. Details on the definition can be found in sec.~\ref{sec:gf}.
The distribution looks reasonably Gaussian with all average values statistically consistent with zero.
We observe for $a=0.079$ fm, that the distribution has slightly larger width, but this is related to the slightly
larger physical volume of that lattice. As we have seen in the previous section, 
the topological susceptibility does not show any sign of discretization errors.

For the computation of the 2-point functions, we have studied 3 different levels $s_i,\,\,i=1,2,3$ of 
Gaussian smearing~\cite{Gusken:1989qx}.
The relevant parameters of the Gaussian smearing that we have considered, usually labeled as  $\{\alpha,N_G\}$,
are $s_1=\{2,30\}$, $s_2=\{4,25\}$, $s_3=\{5.5,70\}$. We have found that the $s_3$ smearing has a better projection
on the fundamental state, but it is also the smearing that adds more noise to the correlator. Compromising 
between an earlier plateau and a less noisier correlator, we have decided to choose the smearing $s_2$ for 
the 2 coarsest lattice spacings and the smearing $s_3$ for the 2 finest spacings.

The fermion lattice action is the non-perturbatively improved 
Wilson action~\cite{Sheikholeslami:1985ij,Luscher:1996sc,Luscher:1996ug}.
The propagators are computed with sources located stochastically in the 3 spatial directions.
We choose $20$ stochastic spatial points for the finest lattice spacing and $10$ stochastic points for the others.
The rational behind this choice is to have O($L/a$) different stochastic points to improve the overlap
between the topological charge and the fermionic part of the correlation functions. We stress that this
is very important to improve the signal-to-noise ratio not only of the 2-point functions, but especially for the 
3-point functions which we discuss in the next section.

We have performed the calculation at $4$ lattice spacings (see sec.~\ref{sec:gf})
and at the following set of momenta
\be
\left\{\mathbbm{P}\right\} = \frac{2 \pi}{L}\cdot\{(0, 0, 0), (\pm 1,0,0),(\pm 1,\pm 1,0),(\pm 1,\pm 1,\pm 1),(\pm 2,0,0)\}\,.
\label{eq:momenta}
\ee
The values of the quark mass for all the lattice spacings corresponds to a value of the 
pseudoscalar mass, $M_{\rm PS}\simeq 800$ MeV, fixed in physical units~\cite{Garden:1999fg}.
From coarser to finer spacings they correspond to the following values of the hopping parameter, 
$\kappa=\left\{0.13353, 0.13423, 0.13460, 0.13485\right\}$.
For these values of $\kappa$ we have computed the nucleon mass, that shows very small discretization
errors and it corresponds to a value $M_N\simeq 1.65$ GeV.
\begin{figure}
\includegraphics[width=16cm]{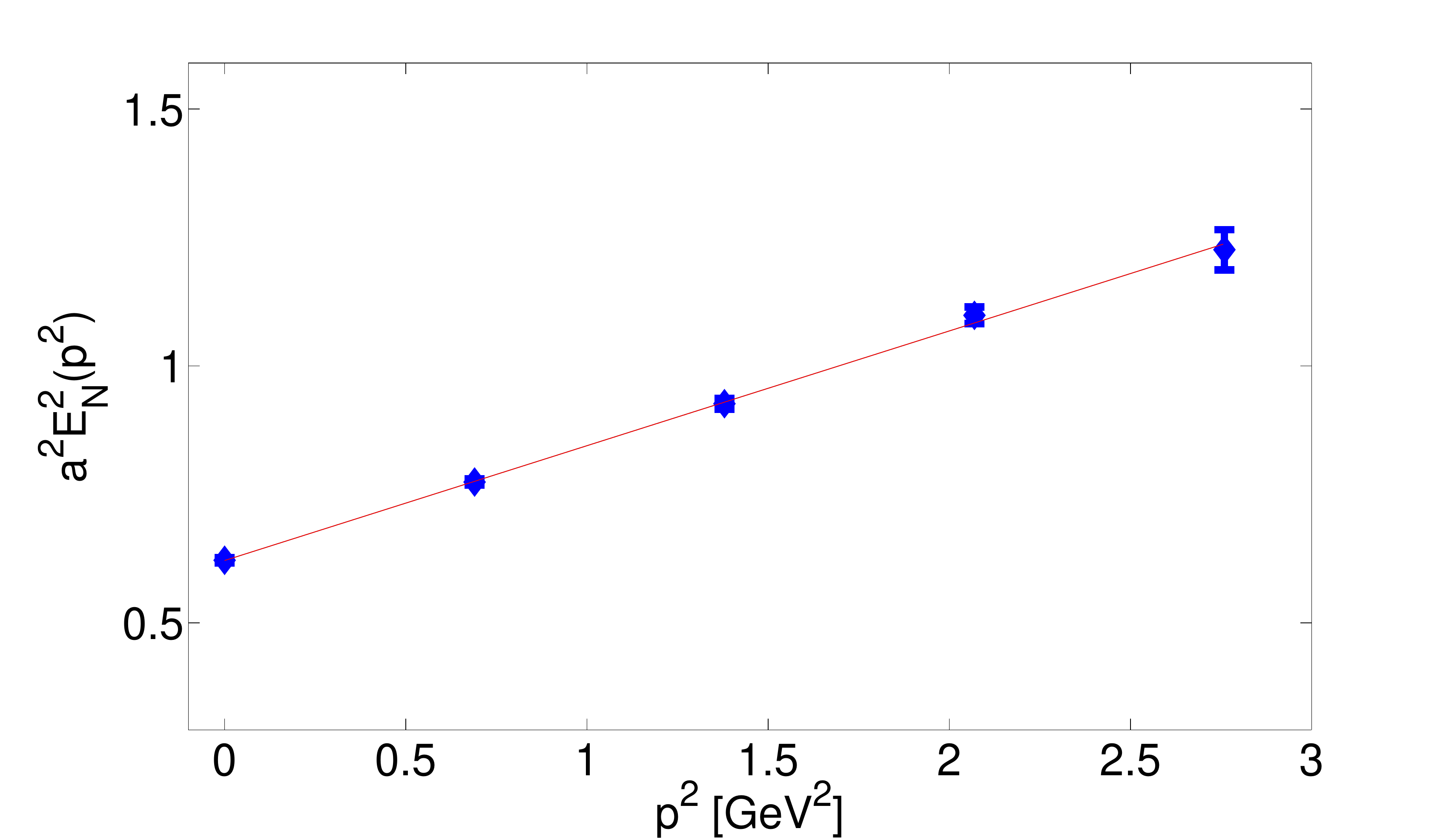}
\caption{Results for the nucleon energy squared at different values of $|\bp|^2$ compared with the continuum dispersion relation
$E_N^2 = M_N^2 + |\bp|^2$.}
\label{fig:dispersion}
\end{figure}
We have also checked the dispersion relation for all lattice spacings and find 
that the discretization errors are below our statistical accuracy. In fig.~\ref{fig:dispersion} we show 
our results for our coarsest lattice spacing of the nucleon energy squared for all the $|\bp|^2$ of the set~\eqref{eq:momenta}
with the continuum form of the dispersion relation. 

\begin{figure}
\includegraphics[width=8cm]{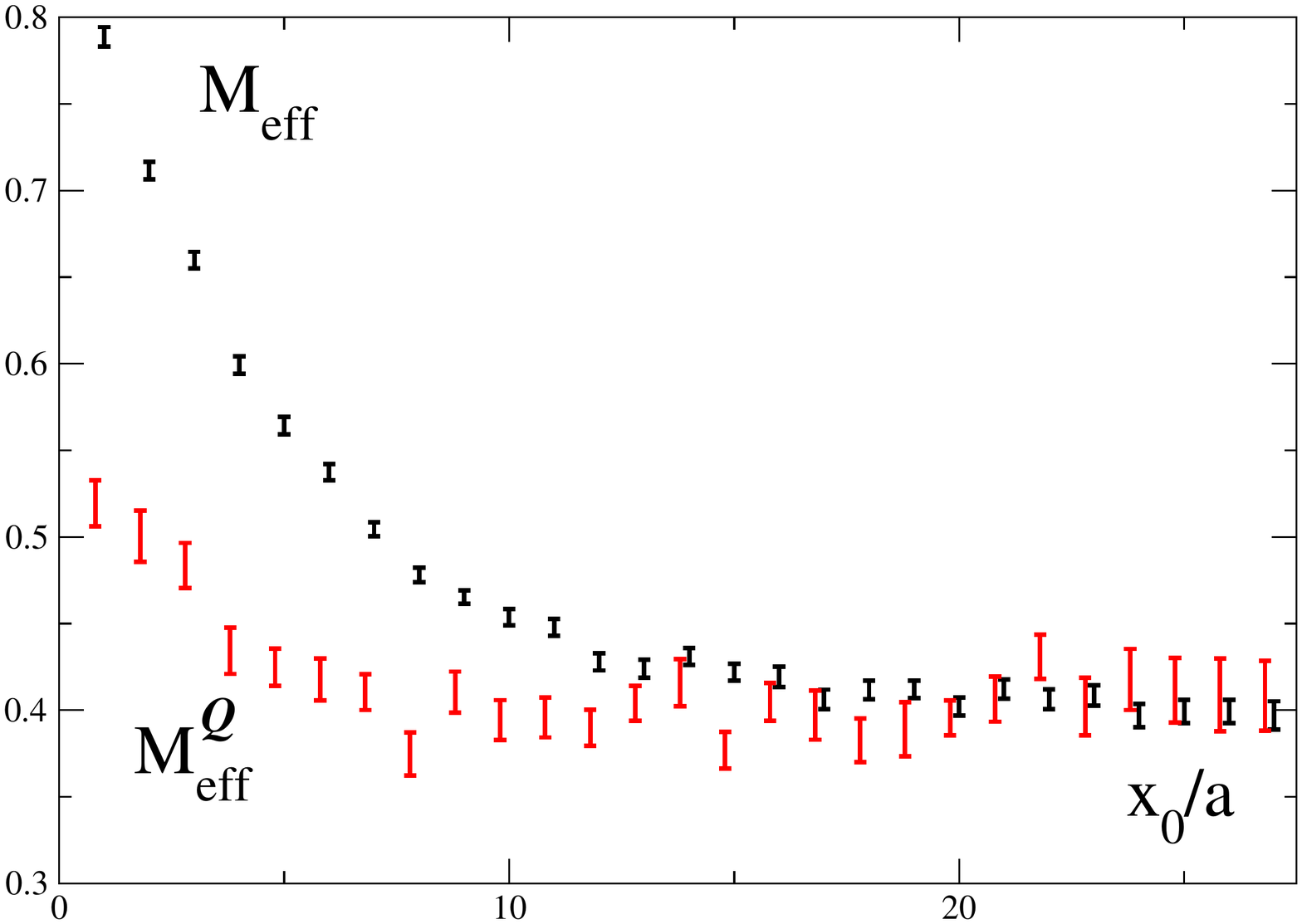}
\includegraphics[width=8cm]{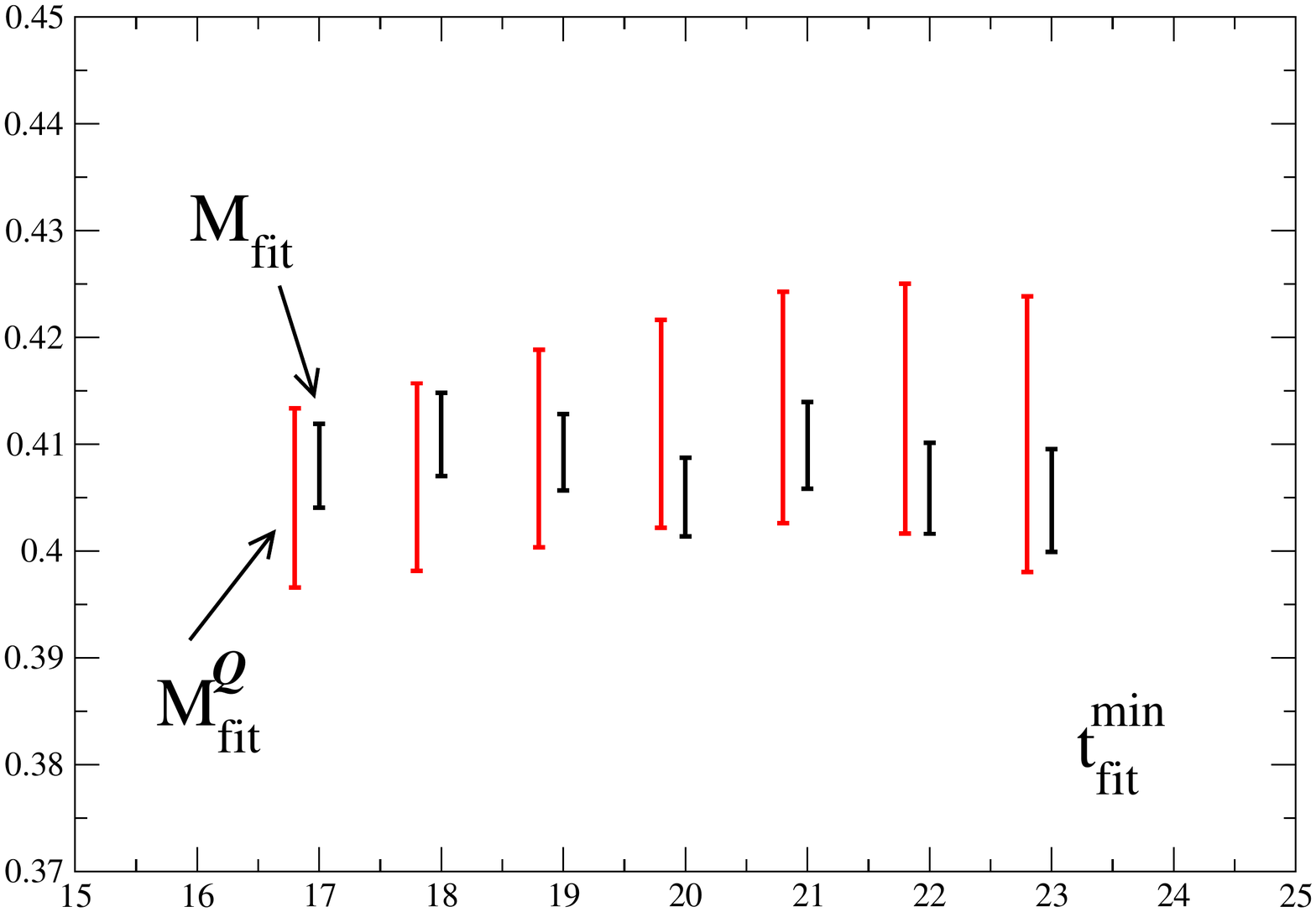}
\caption{Left plot: comparison of the effective masses in lattice units   
obtained from the nucleon correlators at $a=0.048$ fm with and without the insertion of the 
topological charge, i.e. from eq.~\eqref{eq:GNN} ($M_{\rm eff}$) and from eq.~\eqref{eq:GNNQ} ($M_{\rm eff}^{\mcQ}$). 
Right plot: comparison of the nucleon masses in lattice units 
obtained with the nucleon correlators~\eqref{eq:GNN}, $M_{\rm fit}$, and~\eqref{eq:GNNQ}, $M_{\rm fit}^\mcQ$, 
at $a=0.048$ fm for different fit ranges $(t_{\rm fit}^{\rm min}, t_{\rm fit}^{\rm max}=28$).}
\label{fig:eff_mass}
\end{figure}

\begin{figure}
\includegraphics[width=8cm]{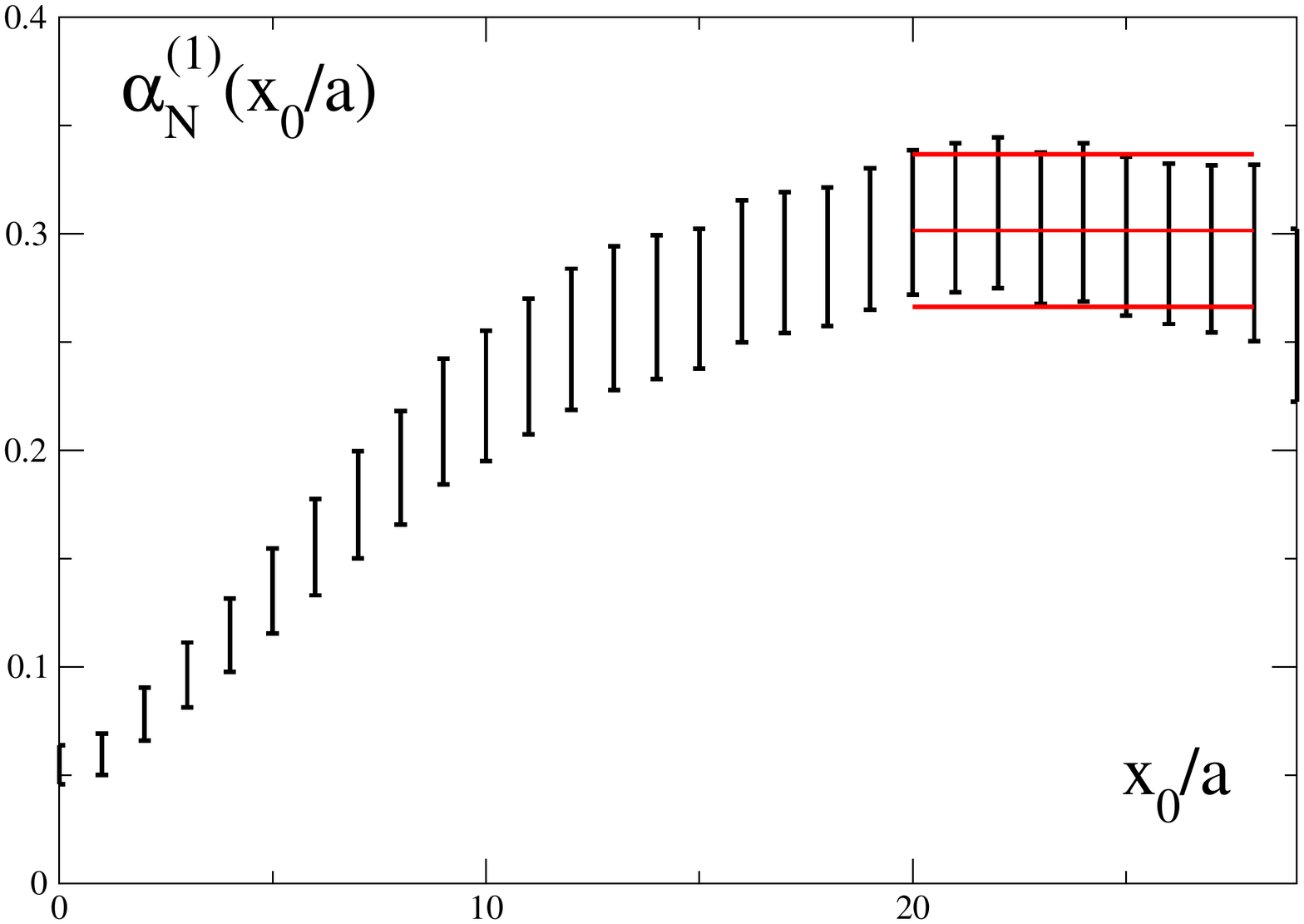}
\includegraphics[width=8cm]{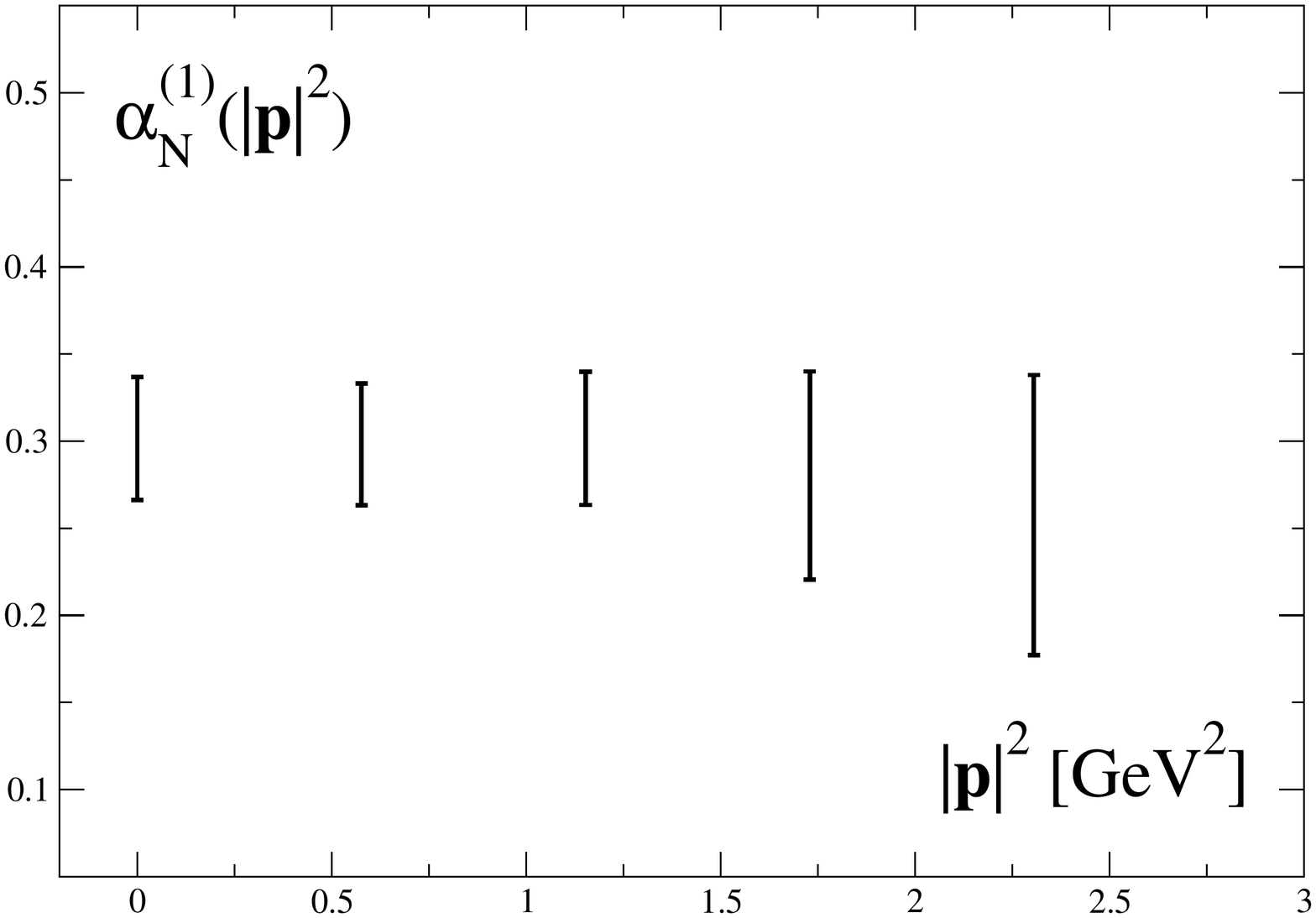}
\caption{Left plot: Euclidean time dependence of $\alpha_N^{(1)}$ determined from the ratio in eq.~\eqref{eq:alpha1}
at $a=0.048$ fm. The red band indicates our choice for the plateau and the corresponding uncertainty. 
Right plot: momentum dependence of $\alpha_N^{(1)}$,
as determined from the ratio in eq.~\eqref{eq:alpha_p}, at $a=0.048$ fm.}
\label{fig:alpha_p}
\end{figure}

In the left plot of fig.~\ref{fig:eff_mass} we show the effective masses of the 
two correlators in eqs.~(\ref{eq:GNN},\ref{eq:GNNQ})
for the finest lattice spacing. It is clear that for large Euclidean times we have perfect
agreement between the two effective masses and very similar results are obtained for all the other lattice
spacings we have. This is also confirmed on the right plot of fig.~\ref{fig:eff_mass} where we show,
again for our finest spacing, the nucleon mass obtained from the 2 correlators in eqs.~(\ref{eq:GNN},\ref{eq:GNNQ})
for different fit ranges.  
The calculation of the mixing angle $\alpha_N^{(1)}$ is now straightforward
\be
\frac{{\rm tr}\left[P_+ \gamma_5 G_{NN}^Q\left(\bzero,x_0\right)\right]}{{\rm tr}\left[P_+G_{NN}\left(\bzero,x_0\right)\right]} = 
\alpha_N^{(1)} + \cdots\,,
\label{eq:alpha1}
\ee
and we expect a plateau for large Euclidean times with higher-order corrections that are exponentially suppressed.
In the left plot of fig.~\ref{fig:alpha_p} we show the Euclidean time dependence of $\alpha_N^{(1)}$ obtained from the
ratio in eq.~\eqref{eq:alpha1} at $a=0.048$ fm. A plateau is easily identified as is the case for all the other lattice spacings.
This is just a reflection of the previous result, namely that asymptotically both correlators in eqs.~(\ref{eq:GNN},\ref{eq:GNNQ}) 
are dominated by the same exponential behavior with the same mass.

We have performed several checks on the calculation of the mixing angle because a solid determination
of $\alpha_N^{(1)}$ is crucial for a correct and precise extraction of the nucleon EDM as detailed in the next section
and in the app.~\ref{app:A}.
We can determine the mixing angle from ratios as in eq.~\eqref{eq:alpha1} but with correlators projected 
at non-vanishing spatial momenta.
If we choose the same interpolating operators for the two correlators, from the spectral 
decomposition in eqs.~(\ref{eq:SDNN},\ref{eq:SDNNQ}) we obtain
\be
\frac{E_N(\bp)+M_N}{2M_N} \frac{{\rm tr}\left[P_+ \gamma_5 G_{NN}^Q(\bp,x_0)\right]}{{\rm tr}\left[P_+G_{NN}(\bp,x_0)\right]} = \alpha_N^{(1)}  + \cdots\,.
\label{eq:alpha_p}
\ee
Up to discretization effects, $\alpha_N^{(1)}$ should not depend on the momentum chosen 
in the nucleon 2-point functions. In the right plot of fig.~\ref{fig:alpha_p} we show the $|\bp|^2$ dependence of $\alpha_N^{(1)}$
for our finest lattice spacing. We expect an increased uncertainty as we increase the nucleon momentum 
and we see perfect agreement between all the values of the mixing angle.
We obtain similar results for all the other lattice spacings.
\begin{figure}
\hspace{-1.0cm}\begin{minipage}{0.5\textwidth}
\centering
\includegraphics[width=9.5cm]{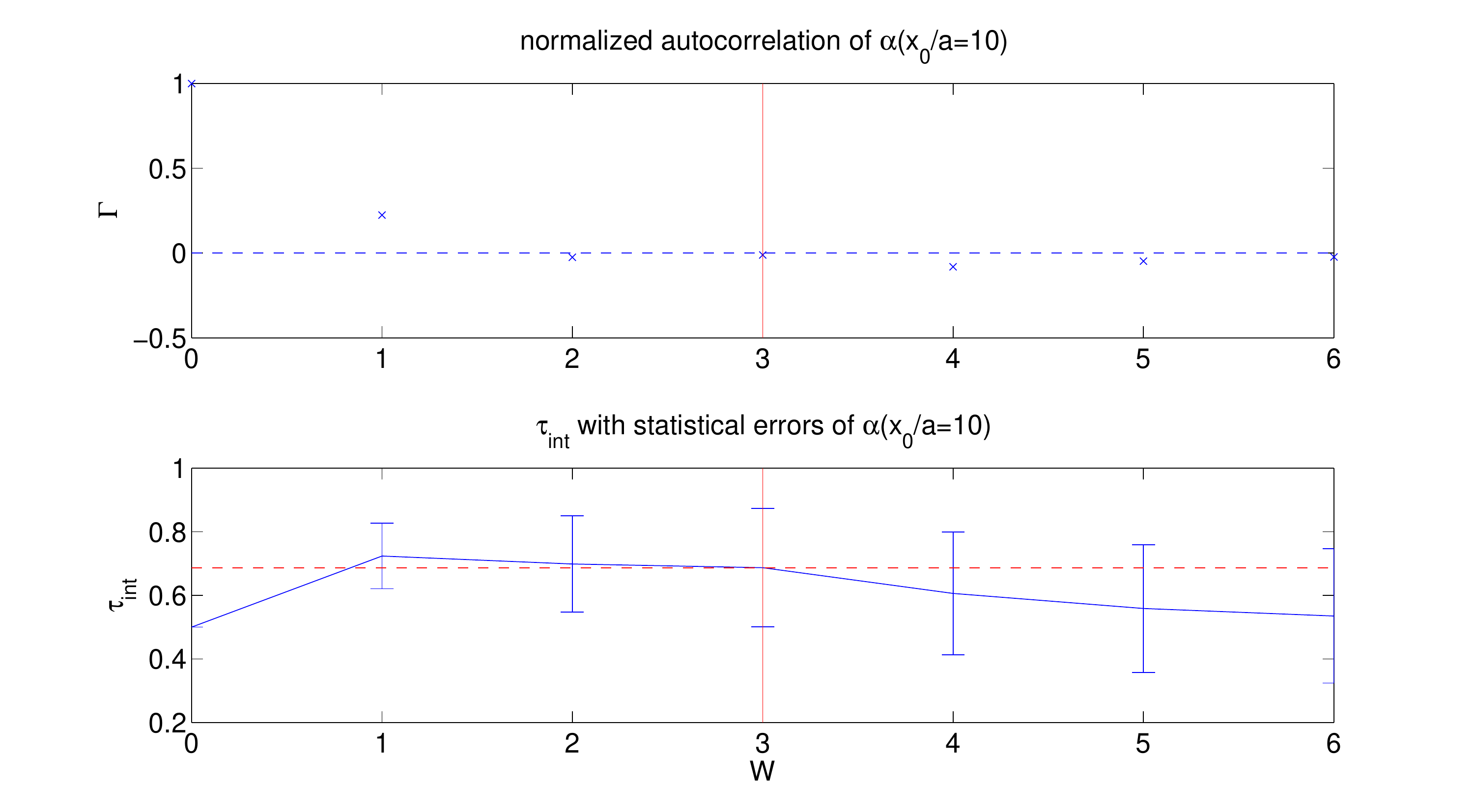}
\end{minipage}
\hspace{1.1cm}\begin{minipage}{0.5\textwidth}
\centering
\includegraphics[width=8cm]{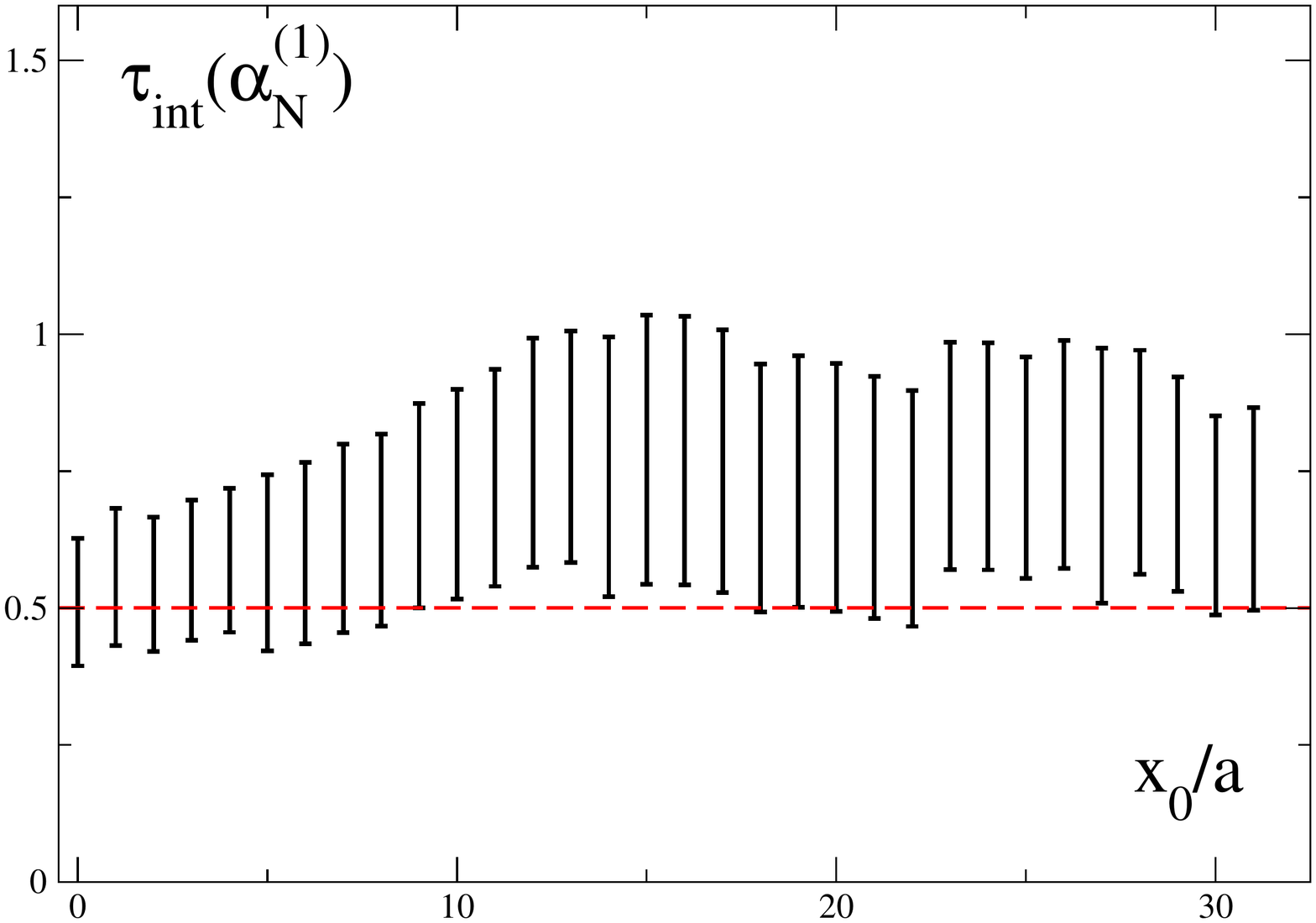}
\end{minipage}
\caption{Left plot: normalized autocorrelation function and estimate of the integrated autocorrelation
time $\tau_{\rm int}$ with the automatic windowing procedure~\cite{Wolff:2003sm} of $\alpha_N^{(1)}$ at $x_0/a=10$. Right plot:
estimate of $\tau_{\rm int}$ of $\alpha_N^{(1)}$ for all Euclidean times. The dashed red line 
indicate the absence of autocorrelation, $\tau_{\rm int}=0.5$.}
\label{fig:tau_alpha}
\end{figure}

Another check of our calculation concerns the autocorrelation time of the correlators containing
the topological charge. Critical slowing down has been observed for topological charge 
and susceptibility both in QCD and Yang-Mills theory~\cite{Schaefer:2010hu} 
using an Hybrid Montecarlo (HMC) algorithm. In particular, it is expected that 
the problem can become relevant for lattice spacings below $0.05$ fm. 
Even though the situation here is different because the correlators contain explicitly fermionic
propagators and we do not use an HMC algorithm, at our finest lattice spacing $a=0.048$ fm we have 
computed the autocorrelation function for all the Euclidean times $x_0$
of the correlator $G_{NN}^\mcQ(\bzero,x_0)$ and $\alpha_N^{(1)}$.
We have followed refs.~\cite{Wolff:2003sm,Madras:1988ei} for the determination of the autocorrelation 
function and integrated autocorrelation times.

\begin{figure}
\centering
\includegraphics[width=11cm]{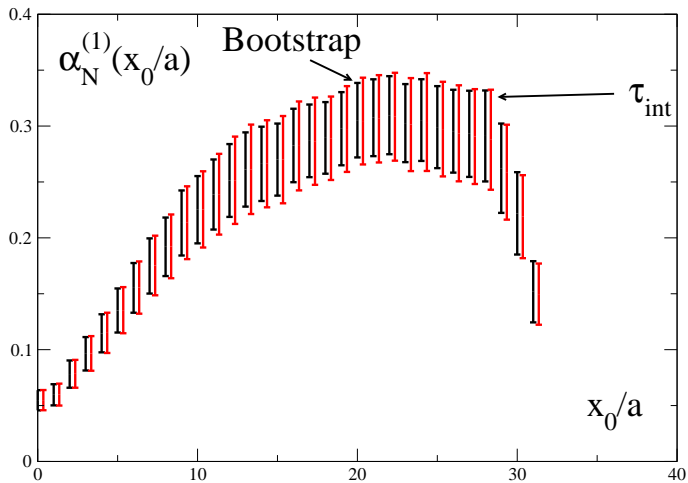}
\caption{Comparison of error estimates using the autocorrelation function method, labeled by $\tau_{\rm int}$,
and a standard bootstrap method.}
\label{fig:alpha_boot_tau}
\end{figure}

\begin{figure}
\includegraphics[width=14cm]{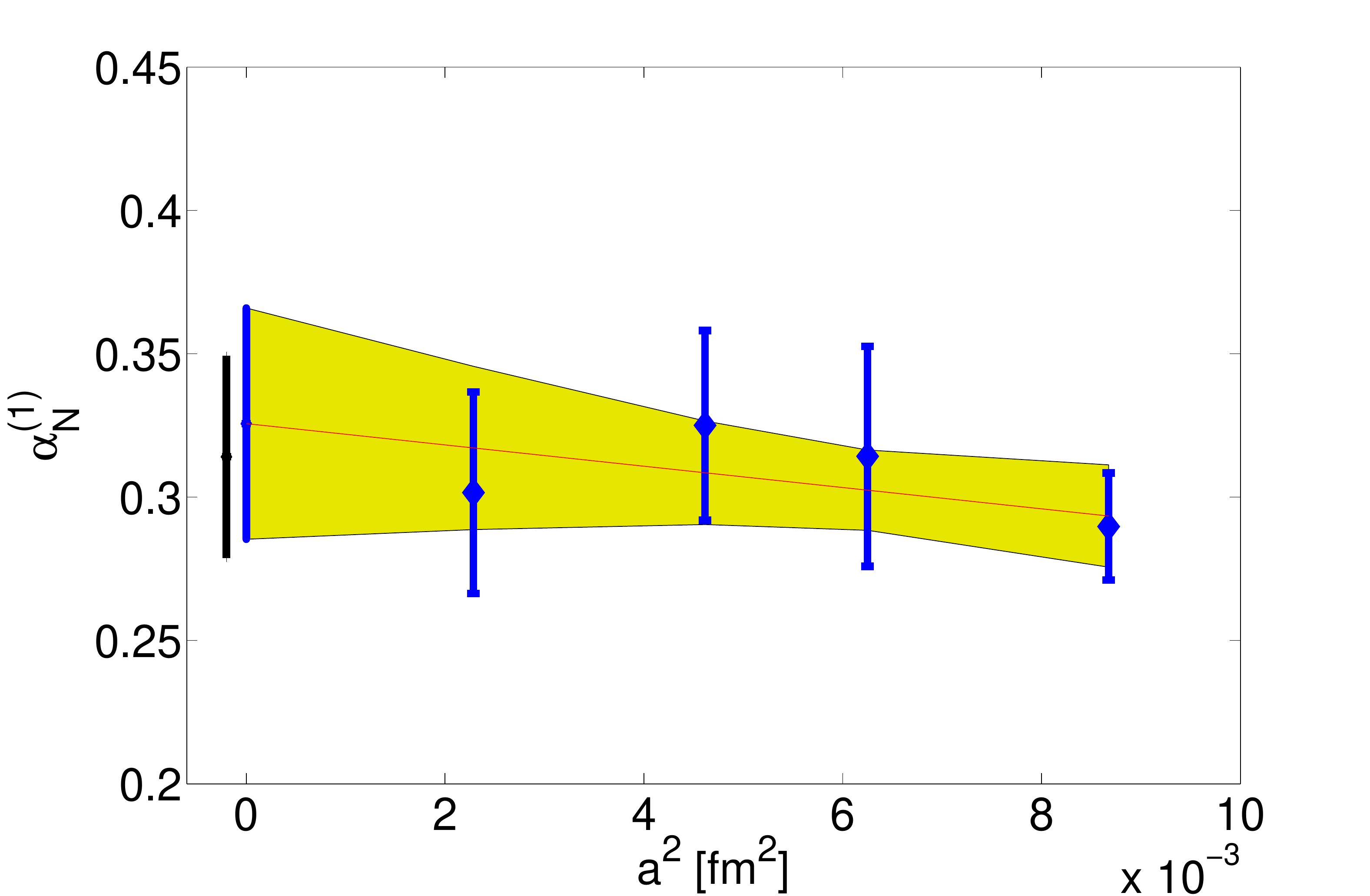}
\caption{Continuum limit of $\alpha_N^{(1)}$. The yellow band is a linear extrapolation 
in $a^2$ and it is compared with a constant extrapolation including all the lattice spacings
and excluding the coarsest one.}
\label{fig:alpha_cl}
\end{figure}

In the left plot of fig.~\ref{fig:tau_alpha} we show for $\alpha_N^{(1)}$ at $x_0/a=10$ 
the normalized autocorrelation function and the estimate of the integrated autocorrelation
time $\tau_{\rm int}$ with an automatic windowing procedure~\cite{Wolff:2003sm}. 
On the right plot we show the estimate of $\tau_{\rm int}$ of $\alpha_N^{(1)}$ for all Euclidean times.
It is clear there is almost no autocorrelation, $\tau_{\rm int}=0.5$, for all Euclidean times.
As a further check, we compared the error estimate of $\alpha_N^{(1)}$ using the autocorrelation function method
to a standard bootstrap method. This is shown in fig.~\ref{fig:alpha_boot_tau} from which
it becomes clear that we can safely use a boostrap analysis 
to determine our statistical uncertainty for all correlators at all our lattice spacings.
We can now perform the continuum limit of $\alpha_N^{(1)}$ for fixed value of the pion mass.
In fig.~\ref{fig:alpha_cl} we show the continuum limit and in tab.~\ref{tab:cl_chi_alpha}
we list all the values at all lattice spacings and different extrapolations to the continuum limit. We compare 
a linear extrapolation in $a^2$ (yellow band) with a constant extrapolation including the three finest lattice spacings. 
We observe a perfect agreement for all the extrapolations and tiny discretization errors. 
The theory is non-perturbatively improved so we expect an O($a^2$) scaling behavior.

Since we see no signs of discretization errors, 
as a final result we quote the value obtained using a constant fit excluding the coarsest lattice spacing
\be
\alpha_N^{(1)}=0.314(35)\,.
\ee
We stress that this is the first time that a continuum limit is performed for this CP-mixing angle.
The normalization chosen for $\alpha_N^{(1)}$ and the convention for the Dirac $\gamma$ matrices
is consistent with the one of ref.~\cite{Berruto:2005hg}.
Our result in the continuum limit differs by $2 \sigma$ from the result of ref.~\cite{Berruto:2005hg}
that is obtained with a different fermionic and gauge action, at a single lattice spacing of
$a\simeq 0.15 $ fm and at a similar quark mass value.

\section{Nucleon electric dipole moment}
\label{sec:nedm}

The spectral decomposition of the 3-point functions 
\be
G_{N J_\mu N}^\theta(\bp_1,\bp_2,x_0,y_0) = a^6 \sum_{\bx,\by} \e^{i\bp_2(\bx-\by)}\e^{i\bp_1\by} \left\langle  \mcN(\bx,x_0) J_\mu(\by,y_0) \bar{\mcN}(0) \right\rangle_\theta\,,
\ee
relevant for the determination of the nucleon EDM is detailed in app.~\ref{app:A} and the final result
for the leading exponentials is given in eq.~\eqref{eq:G3t_SD}.
By taking suitable ratios of 2- and 3-point functions we can extract CP-even and CP-odd form
factors defined in eqs.~(\ref{eq:gamma_evenQ2},\ref{eq:gamma_oddQ2}). 

The 3-point functions have been computed for the set of momenta $\left\{\mathbbm{P}\right\}$ defined in 
eq.~\eqref{eq:momenta} and, when possible, we have averaged over all equivalent momenta configurations.
We have tested several sink locations $x_0$ and, after some numerical experiments,
we have chosen the following set 
\be
\left\{x_0\right\} = (16 a, 20 a, 20 a, 28 a)\,,
\ee
from the coarsest to the finest lattice spacing.
We work in the SU($3$) flavor-symmetric limit such that the disconnected contributions vanish.

\begin{figure}
\includegraphics[width=8cm]{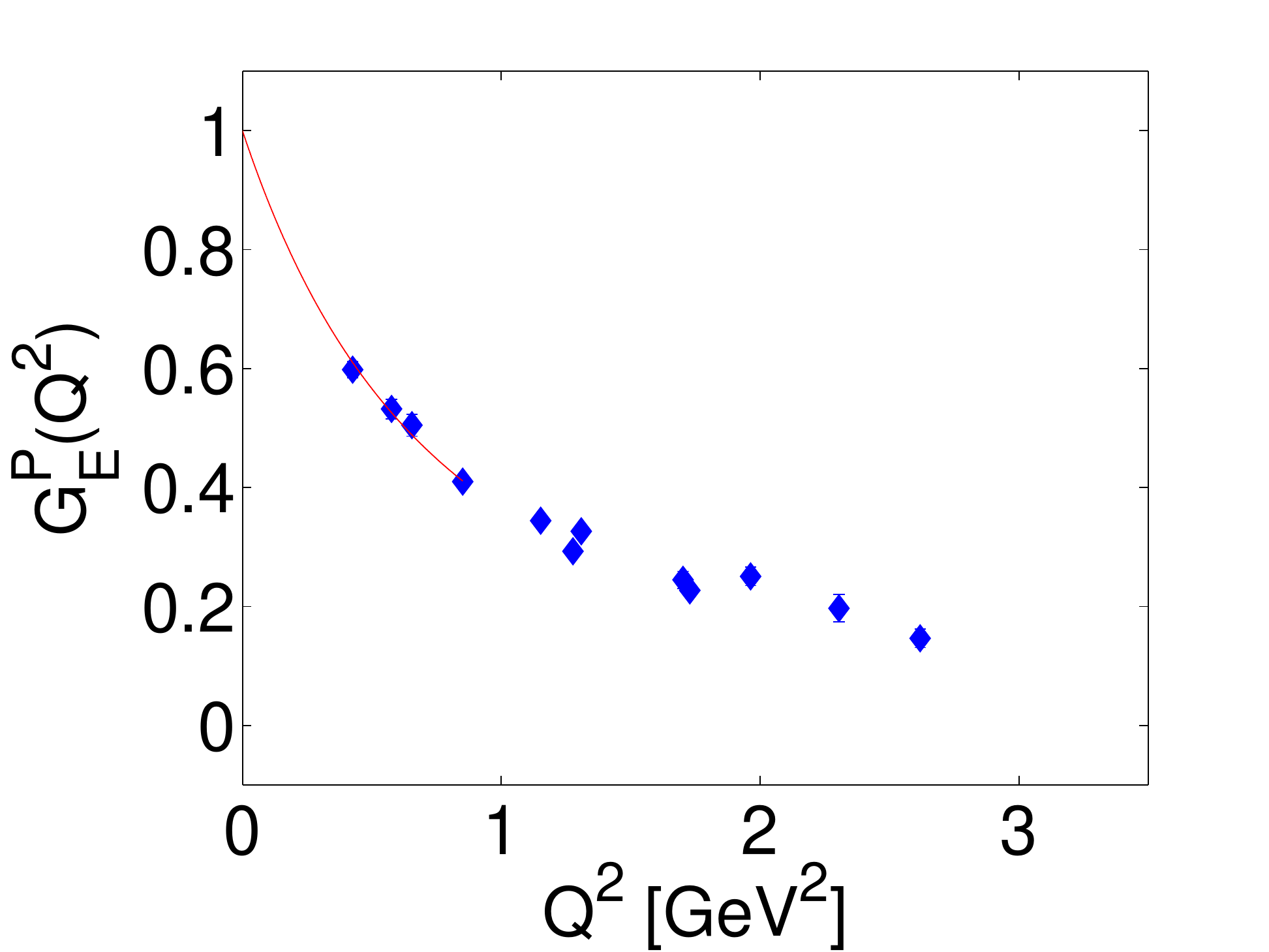}
\includegraphics[width=8cm]{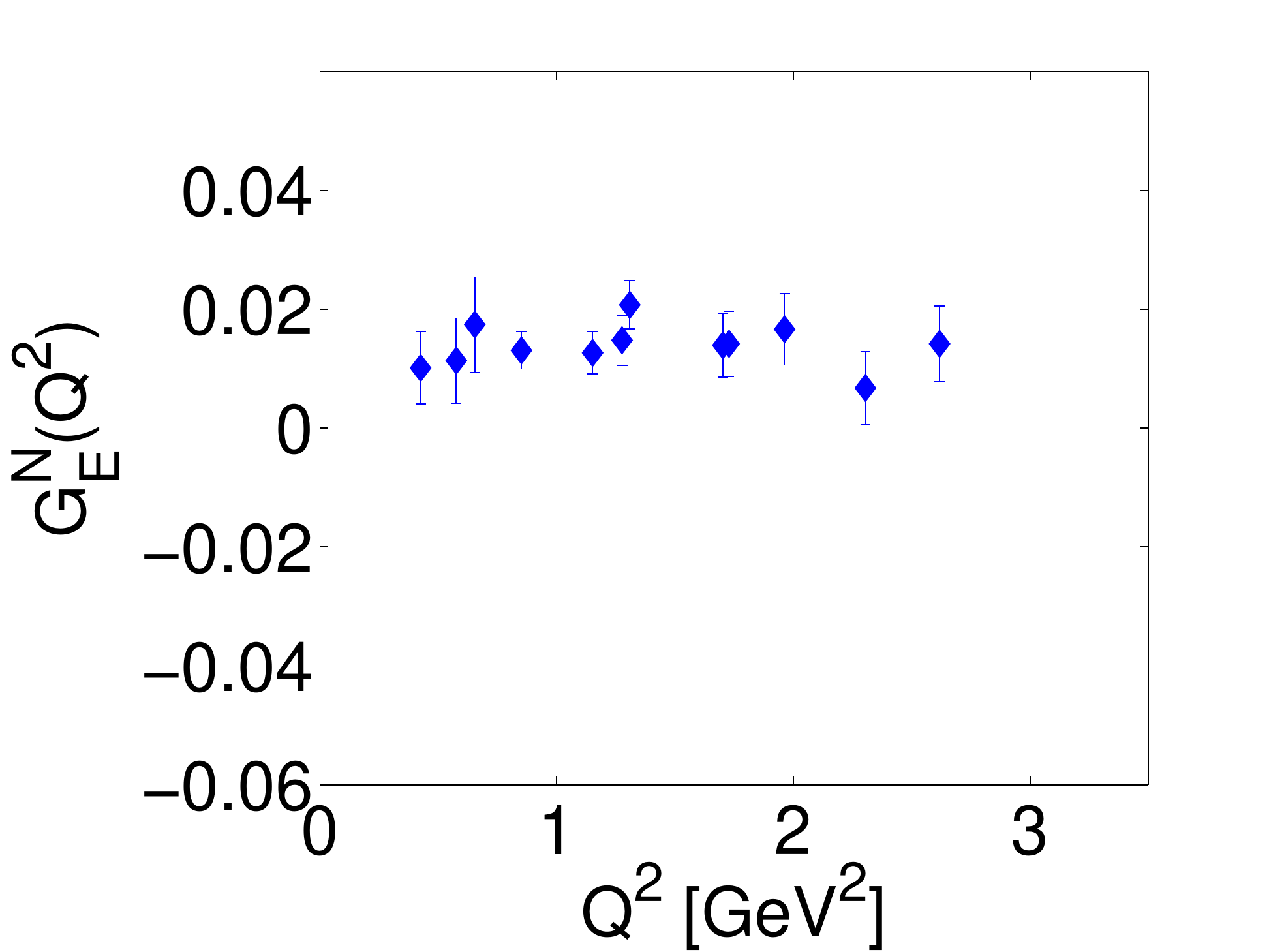}
\caption{Momentum dependence of the electric form factor for the proton (left plot) and the neutron (right plot).
The red curve for the proton form factor is a phenomenological fit restricted to the 4 lowest momenta. The fit result is
$M_D^2 = 1.5182(71)~{\rm GeV}^2$ (see text).}
\label{fig:ge}
\end{figure}

\begin{figure}
\includegraphics[width=8cm]{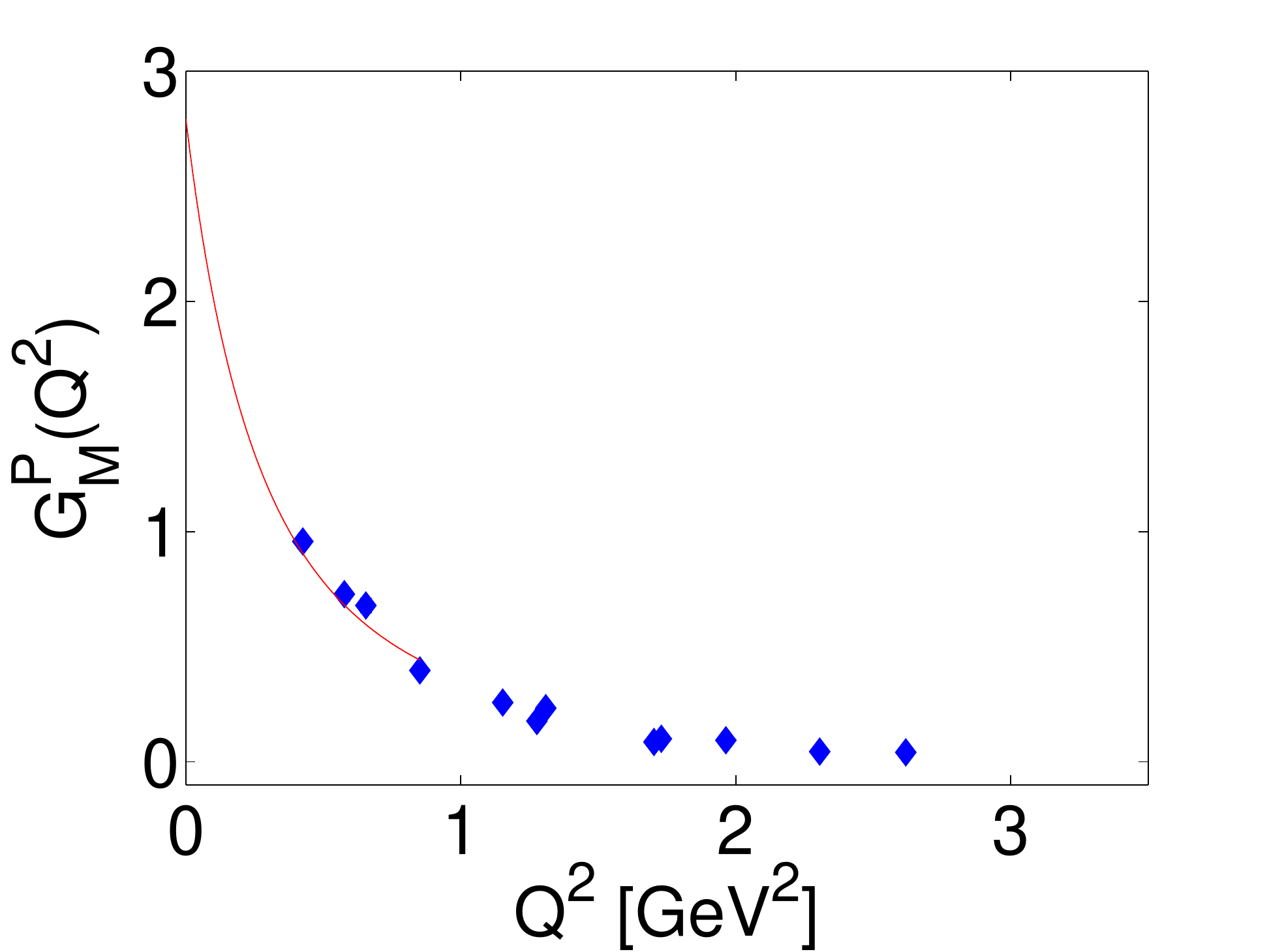}
\includegraphics[width=8cm]{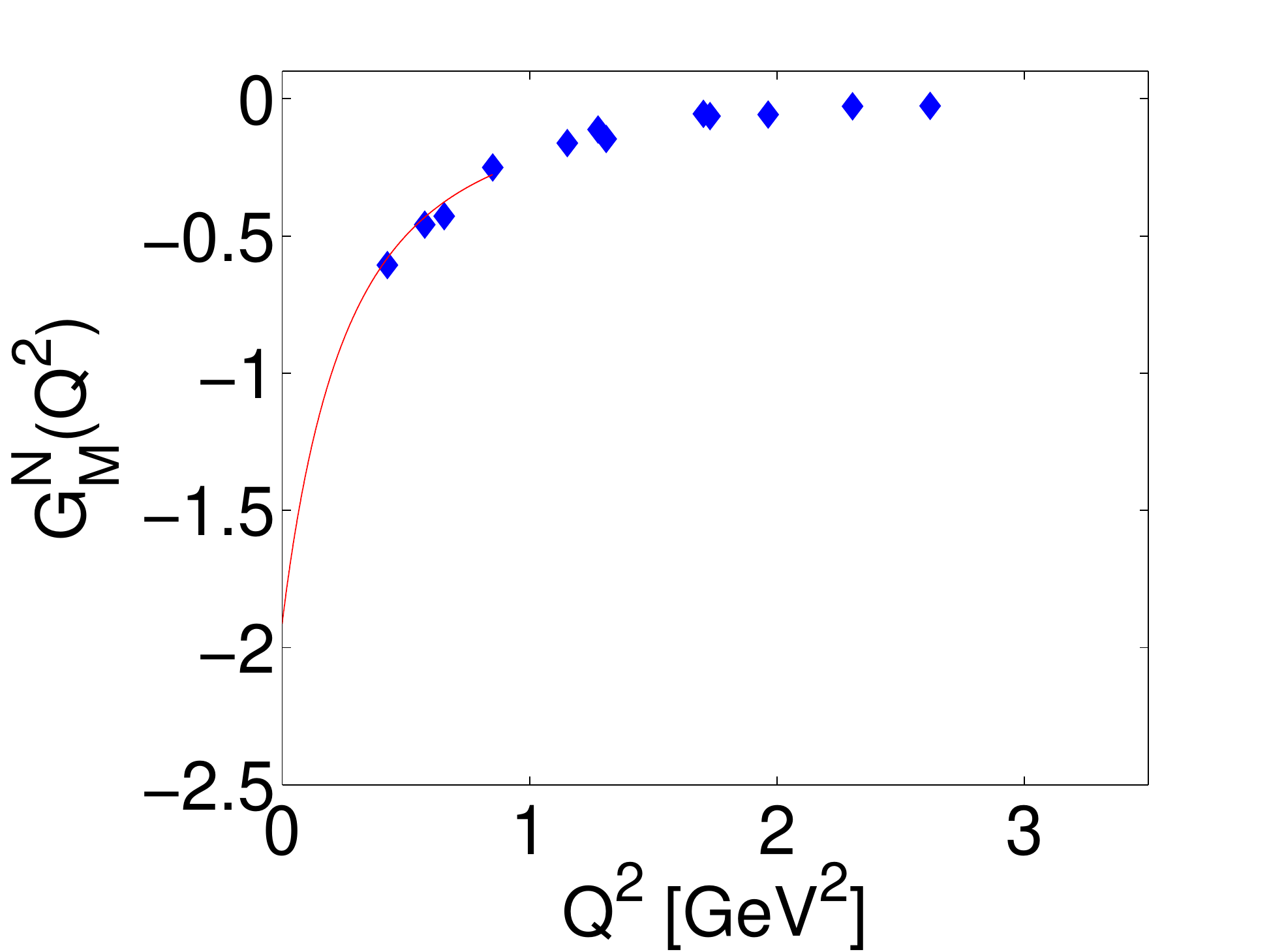}
\caption{Same as fig.~\ref{fig:ge} but for the magnetic form factors. The fit results are 
$M_D^2=0.5621(20)~{\rm GeV}^2$ for the proton and $M_D^2=0.5224(15)~{\rm GeV}^2$ for the neutron.}
\label{fig:gm}
\end{figure}

For all the form factors calculations we use a local vector current. The normalization constant
$Z_V(g_0^2)$ is taken from ref.~\cite{Luscher:1996jn}. 
To compute the EDM we use the ratio of eq.~\eqref{eq:R05}. In order to determine $F_3(Q^2)$, 
we need to subtract contributions proportional to the mixing angle $\alpha_N^{(1)}$ and the CP-even form factors
$G_E(Q^2)$ and $G_M(Q^2)$.
\begin{figure}
\includegraphics[width=16cm]{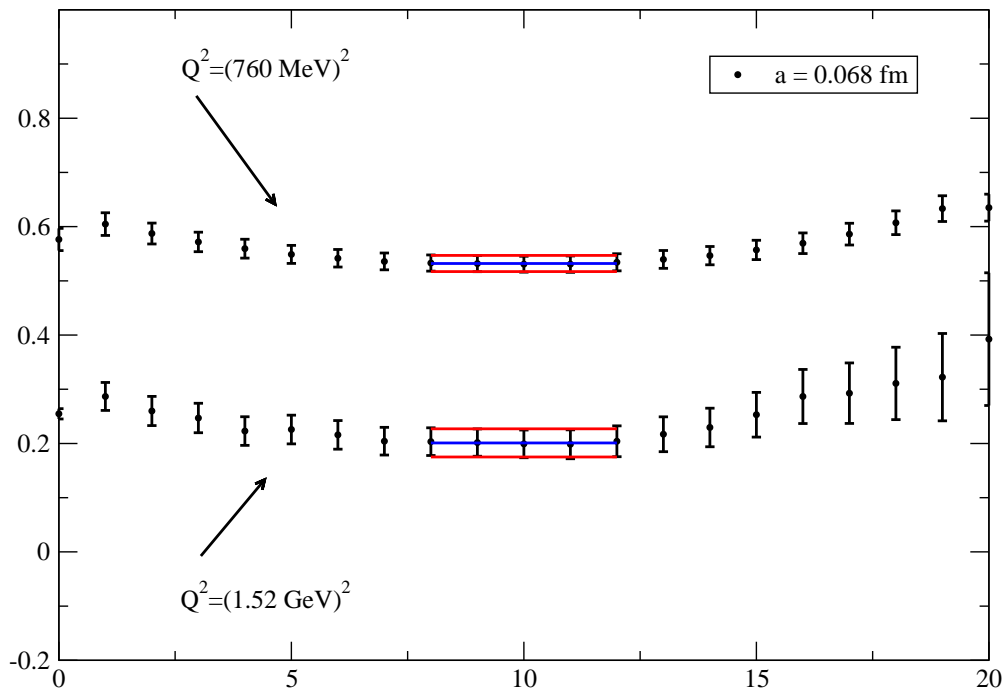}
\caption{Euclidean time dependence of the proton electric form factor determined from the ratio~\eqref{eq:R0b} for 2 different values of $Q^2$.}
\label{fig:gep_pl}
\end{figure}
To determine $G_E(Q^2)$ and $G_M(Q^2)$ we use the ratios in eqs.~\eqref{eq:R0b} and \eqref{eq:Ri}.
In fig.~\ref{fig:gep_pl} we show 2 typical plateaus for the electric form factor of the proton for 2 different
momenta. A plateau is easily identified as for all the other 
momenta and lattice spacings. We collect in a single plot, fig.~\ref{fig:ge}, the electric form factor $G_E(Q^2)$ of the proton 
and the neutron including all the lattice spacings. Discretization errors are well under control.
The red curve for the proton electric form factor is a fit to the lowest 4 momenta using the standard dipole form
\be
G_E^{\rm dip}(Q^2) = \frac{1}{\left(1+\frac{Q^2}{M_D^2}\right)^2}\,.
\ee
The neutron electric form factor vanishes at $Q^2=0$ and is rather small for larger values of $Q^2$,
but we are still able to identify a clear signal over a wide range of $Q^2$. 
This will be important for the determination of the neutron EDM.
In fig.~\ref{fig:gm} we show the same result for the magnetic form factors and the same type of dipole fit 
\be
G_E^{\rm dip}(Q^2) = \frac{\kappa_{P,N}}{\left(1+\frac{Q^2}{M_D^2}\right)^2}\,,
\ee
where the anomalous magnetic moments $\kappa_{P,N}$ are fixed to their phenomenological
values. The fit parameters $M_D^2$ are not expected to reproduce the phenomenological values 
(see for example ref.~\cite{Belushkin:2006qa}).
We perform these fits as a check that our lattice data can be fitted by a dipole form, 
however we do not use these fits in the determination of the EDMs below. 

\begin{figure}
\includegraphics[width=16cm]{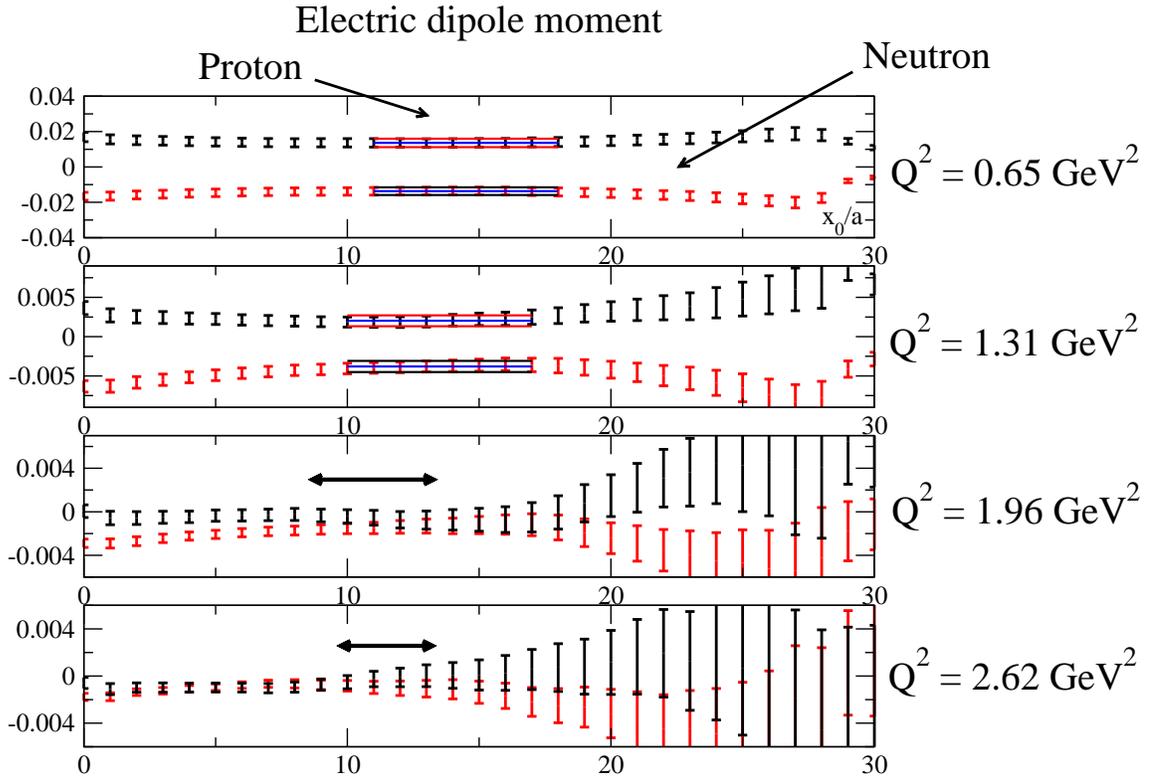}
\caption{Euclidean time dependence of $F_3(Q^2)/2M_N$ in $e~\cdot$ fm extracted from the ratio in eq.~\eqref{eq:R05}
for the 4 non-vanishing lattice momenta at our finest lattice spacing.}
\label{fig:edm_pl}
\end{figure}

\begin{figure}
\includegraphics[width=16cm]{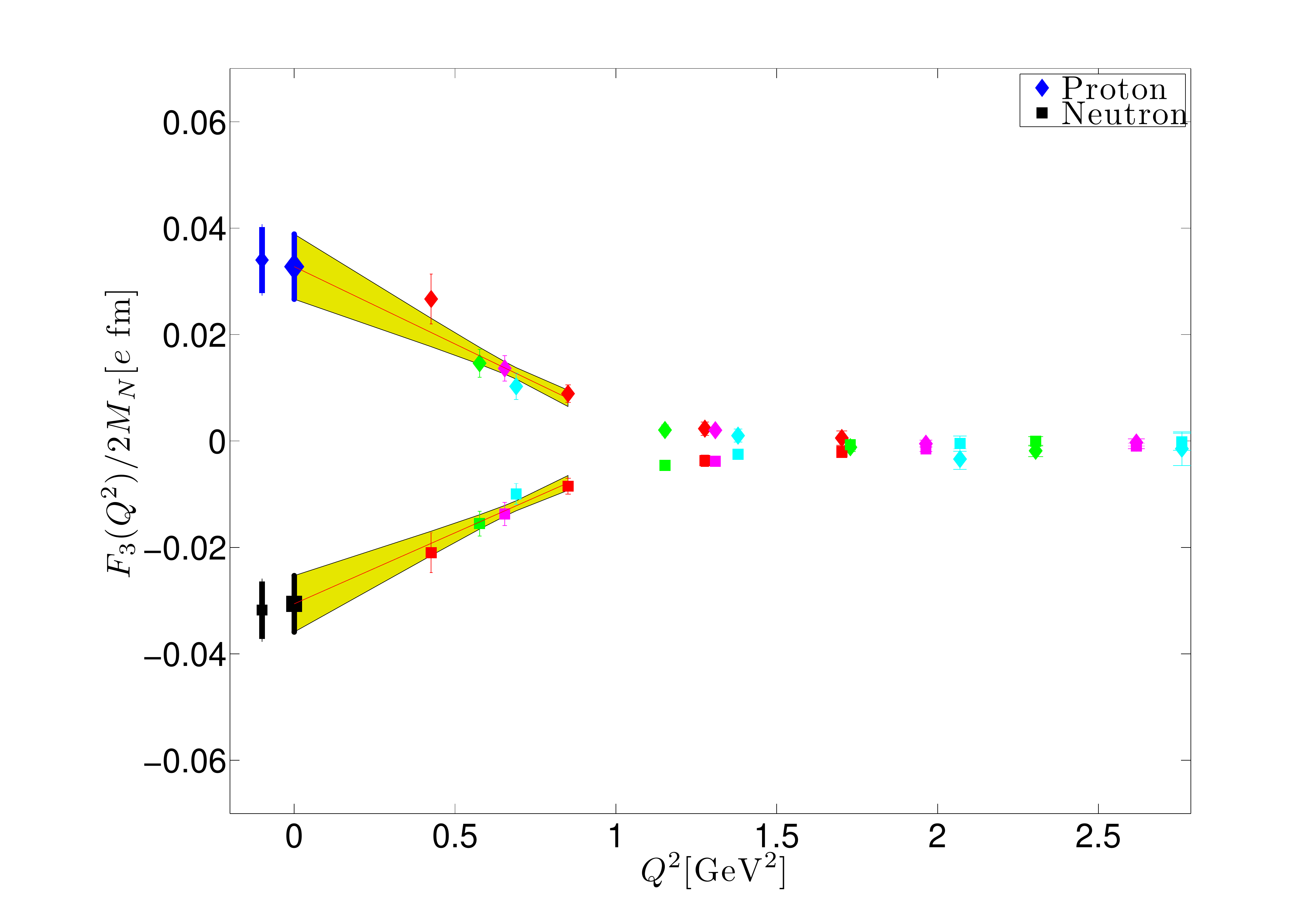}
\caption{Momentum dependence of the CP-odd form factor $F_3(Q^2)/2M_N$ of the proton and the neutron. The yellow
band is a linear extrapolation in $Q^2$ as suggested from $\chi$PT at NLO using all four lattice
spacings. Different colors represent different
lattice spacings: $\beta=6.0$ (cyan), $\beta=6.1$ (red), $\beta=6.2$ (magenta), $\beta=6.45$ (green).
As a comparison we plot, slightly displaced, the $Q^2=0$ extrapolation using the three finest lattice spacings.}
\label{fig:edm_Q2}
\end{figure}

With the precise determination of the CP-even form factors and the mixing angle $\alpha_N^{(1)}$ 
we can now determine the nucleon EDM. 
By evaluating the ratio on the l.h.s. of eq.~\eqref{eq:R05} and subtracting 
the spurious contributions from the r.h.s, we can determine the CP-odd form factor $F_3(Q^2)$.
In fig.~\ref{fig:edm_pl} we show the plateau for the normalized CP-odd form factor 
$F_3(Q^2)/2M_N$ for all the momenta at our finest lattice spacing.
For the lowest 2 momenta it is possible to extract a signal while for the largest 2, 
the signal is too small and consistent with zero.
In fig.~\ref{fig:edm_Q2} we show the $Q^2$ dependence of $F_3(Q^2)/2M_N$ on a single plot including
all our lattice spacings. The different lattice spacings results agree rather well, 
indicating small discretization errors within our statistical accuracy.
For this reason, we determine the EDM by extrapolating using the three finest lattice spacings result to $Q^2=0$ 
with the same fit function. An extrapolation using all four lattice spacings give completely consistent results
as shown in fig.~\ref{fig:edm_Q2}.
For the extrapolation to $Q^2=0$, we use the SU($2$) $\chi$PT result of~\cite{Mereghetti:2010kp} as a guideline.
There the form factor is expanded as 
\be
\frac{F_3^{P/N}(Q^2)}{2 M_N} = d_{P/N} + S_{P/N} Q^2 + H_{P/N}(Q^2)\,.
\ee
The values at $Q^2=0$ are the nucleons EDMs and the slope in $Q^2$ at small $Q^2$, $S_{P/N}$, 
are the so called Schiff moments~\cite{Thomas:1994wi}.
The functions $H_{P/N}(Q^2)$, defined in~\cite{Mereghetti:2010kp}, scale as $Q^4$
for small $Q^2$ and they can be neglected for small enough values of $Q^2$. 
The numerical data for small $Q^2$ indeed suggest a linear $Q^2$ dependence.
A linear extrapolation in $Q^2$, using the three finest lattice spacings, gives us the values of the proton and neutron EDMs
\be
d_{\rm P} = 0.0340(62) ~ \theta ~e \cdot {\rm fm}\,,
\label{eq:dP}
\ee
\be
d_{\rm N} = -0.0318(54) ~ \theta ~e \cdot {\rm fm}\,.
\label{eq:dN}
\ee
If we make the reasonable assumption that at this relatively large value of the 
pseudoscalar mass, quenched and unquenched calculations give comparable results,
we can try to estimate the values of the EDM at the physical point. To do so, we use
as constraint the fact that the EDM in the continuum
has to vanish in the chiral limit.
In principle, we would like to use the $\chi$PT expressions in eqs. 
\eqref{eq:pEDM} and \eqref{eq:nEDM}, but considering the large pseudoscalar mass used
in our calculations, the $\chi$PT expressions are not reliable and we instead perform a simple linear fit in $M_\pi^2$. 
We then obtain the following estimates
\be
d_{\rm P}^{\rm phys} = 0.96(18) \cdot 10^{-3}~ \theta ~e \cdot {\rm fm}\,,
\ee
\be
d_{\rm N}^{\rm phys} = -0.90(15) \cdot 10^{-3} ~ \theta ~e \cdot {\rm fm}\,,
\ee
where we have only included the errors from eqs.~\eqref{eq:dP} and \eqref{eq:dN}.
These estimates are statistically consistent with the results of~\cite{Guo:2012vf,Akan:2014yha}.
We stress that many systematic uncertainties are not taken into account in our calculation and that
the main goal of this work is to describe the new methodology and perform a first continuum
extrapolation.
With all the caveats intrinsic in our calculation, we can extract an upper bound for $\theta$.
The experimental upper bound of the neutron EDM, $ |d_N| < 2.9 \cdot 10^{-13} e \cdot$ fm, gives
$\theta \lesssim 3.2 \cdot 10^{-10}$. 

Our lattice data also allows us to extract the nucleon Schiff moments for which we obtain
\be
S_{\rm P} = -1.16(33) \cdot 10^{-3} ~ \theta ~e \cdot {\rm fm^3}\,,
\ee
\be
S_{\rm N} = 1.07(28) \cdot 10^{-3} ~ \theta ~e \cdot {\rm fm^3}\,.
\ee
In both $SU(2)$ \cite{Hockings:2005cn,Mereghetti:2010kp} and $SU(3)$ \cite{Ottnad:2009jw,deVries:2015una} 
$\chi$PT, the nucleon Schiff moments are of isovector nature, 
in agreement with our lattice results. 
In fact, at leading order in the $SU(2)$ chiral expansion, the Schiff moments are predicted
\be
S_{\rm P} = -S_{\rm N} = - \frac{e g_A \bar{g}_0^\theta}{48 \pi^2 F_\pi^2 M_\pi^2} = 
-(1.8\pm 0.2 )\cdot 10^{-4}\,\theta\,e\,\cdot \mathrm{fm}^3\,\,\,,
\ee
where we used the value of $\bar{g}_0^\theta$ given in eq.~(\ref{eq:g0theta}). 

We see that our lattice results are roughly $5$ times larger than the leading-order
$\chi$PT predictions which are in principle pion-mass independent (note that $\bar{g}_0^\theta \sim M_\pi^2$). 
However, already at the physical pion mass, the nucleon Schiff moments obtain $\mathcal O(60\%)$ 
next-to-leading-order corrections that scale as $\sim M_\pi$ \cite{Mereghetti:2010kp,Ottnad:2009jw}. 
Considering the large pseudoscalar mass used in our calculation, the discrepancy is not very worrisome.  

To conclude, we have shown that it is possible to obtain a non-perturbative determination of the Schiff moments from 
the $Q^2$ dependence of the CP-odd form factor, $F_3$. 
With more precise lattice data at smaller pion masses, the extraction of the nucleon Schiff moments 
can be used for a direct determination of the LEC $\bar{g}_0$. 
In this way, from the pion mass dependence of the EDMs (see eqs.~(\ref{eq:nEDM},\ref{eq:pEDM})), 
it is then possible to extract the other LECs, $\Lambda_{P,N}$ (or equivalently $\bar d_{N,P}$), 
from lattice QCD alone without relying on eq.~(\ref{eq:g0theta}).

\section{Final remarks}
\label{sec:conclu}

We have presented a first-of-its-kind continuum limit for the CP-mixing angle $\alpha_N$ and
the nucleon EDMs. At the same time, we have performed a first \textit{ab initio} 
calculation of the nucleon Schiff moments in the continuum limit which can be used 
to extract the value of the CP-odd pion-nucleon coupling constant $\bar g_0$.
The key ingredient is the use of the gradient flow for the definition
of the topological charge which, in this way, 
is free of renormalization ambiguities and allows a straightforward
continuum extrapolation. The method we have proposed is general and can be used
for any quantity computed in a $\theta$ vacuum such as EDMs of light nuclei.
To test this new method we have performed the calculation in the Yang-Mills theory at
a relatively large value of the quark mass. We are currently extending this calculation
to QCD with dynamical configurations and smaller quark masses.

Previous calculations of nucleon EDMs~\cite{Shintani:2005xg,Berruto:2005hg},
also applied a perturbative expansion in $\theta$, 
but instead used a cooling procedure to define the topological charge.
We do believe that such calculations give the right qualitative answer, but 
we stress that defining the EDM in this way does not allow for a controlled
continuum limit.
Another method that has been proposed is to consider 
an imaginary $\theta$-term in the action~\cite{Guo:2015tla}.
With an imaginary $\theta$ the action becomes real and amenable to numerical
Monte Carlo methods, but it requires the generation of a new gauge ensemble for each value of $\theta$.
The range of $\theta$ used for the generation of these ensembles is $\theta \simeq 0.5-2.5$.
If the calculation is performed with a Wilson-type
action, the $\theta$ coefficient needs to be renormalized in order to restore the
proper anomalous Ward identity. Additionally, the analytic continuation back to a real
value of $\theta$ has to be done with care in regions outside the perturbative regime in $\theta$.
These complications are avoided completely with our proposal, 
because we directly compute the linear coefficient of $\theta$ in the 
standard QCD background.

As a last remark we recall that with a Wilson-type fermion action, 
EDMs are not guaranteed to vanish in the chiral limit, which only happens after the 
continuum limit has been performed.
The same phenomenon takes place for the topological susceptibility~\cite{Bruno:2014ova}.
It is only {\it after} performing the continuum limit that it is possible to constrain the 
$\theta$-induced EDM to vanish in the chiral limit. 
This stresses the importance of performing
the continuum limit prior to any chiral limit 
and we believe that our method is optimal in this respect.

We consider this work as a first step in the determination of $\theta$ 
and beyond-the-Standard-Model matrix elements with the gradient flow. 
Other contributions to EDMs, for instance from fermionic operators such as 
quark chromo-electric dipole moments, can be determined using the gradient flow 
for fermions~\cite{Luscher:2013cpa} and work in this direction is in progress.

\section*{Acknowledgments}
We thank R. Edwards and B. Jo\'o for help with CHROMA~\cite{Edwards:2004sx}.
We thank U.-G. Mei{\ss}ner for a careful reading of the manuscript.
This work (JdV) is supported in part by the DFG and the NSFC
through funds provided to the Sino-German CRC 110 ``Symmetries and
the Emergence of Structure in QCD'' (Grant No. 11261130311).
The authors gratefully acknowledge the computing time granted by the JARA-HPC 
Vergabegremium and provided on the JARA-HPC Partition part of the supercomputer 
JUQUEEN at Forschungszentrum J\"ulich.

\vspace{0.4cm}
\begin{appendix}
\section{Appendix A}
\label{app:A}

In this appendix we discuss in some detail the spectral decomposition of the 3-point functions
used to determine the nucleon form factors in a $\theta$ vacuum.
For completeness we remind that in our calculations
the initial momentum is $p_1=((p_1)_0,\bp_1)$ and the final momentum is $p_2=((p_2)_0,\bp_2)$. 
The momentum transfer is $q=p_2-p_1=(E(\bp_2)-E(\bp_1),\bp_2-\bp_1)$.

If we retain only the leading exponential contribution, the spectral decomposition of the 3-point functions
\be
G_{N J_\mu N}^\theta(\bp_1,\bp_2,x_0,y_0) = a^6 \sum_{\bx,\by} \e^{i\bp_1\by} \e^{i\bp_2(\bx-\by)} \left\langle \mcN(\bx,x_0) J_\mu(\by,y_0) 
\overline{\mcN}(0)\right\rangle_\theta\,,
\label{eq:GNJNt}
\ee
is given by
\bea
G_{N J_\mu N}^\theta(\bp_1,\bp_2;x_0,y_0) &=& \frac{{\rm e}^{-E_N(\bp_1)y_0}}{2 E_N(\bp_1)} 
\frac{{\rm e}^{-E_N(\bp_2)(x_0-y_0)}}{2 E_N(\bp_2)} \\ \nonumber
&\times& \sum_{s,s'}\langle \theta |\mcN |N^\theta(\bp_2,s')\rangle 
\langle N^\theta(\bp_2,s')| J_\mu | N^\theta(\bp_1,s)\rangle
\langle N^\theta(\bp_1,s)| \overline{\mcN}| \theta \rangle\,.
\eea
Following the parametrization in eq.~\eqref{eq:gamma_Q2} and using the completeness relation, 
for we obtain for small values of $\theta$
\bea
&\phantom{+}&G_{N J_\mu N}^\theta(\bp_1,\bp_2;x_0,y_0)_{\alpha\beta} = \frac{{\rm e}^{-E_N(\bp_1)y_0}}{2 E_N(\bp_1)} 
\frac{{\rm e}^{-E_N(\bp_2)(x_0-y_0)}}{2 E_N(\bp_2)} \mcZ_N^*(\bp_1) \mcZ_N(\bp_2) \\ \nonumber
&\times& \left\{\left[E_N(\bp_2)\gamma_0 - i \gamma_k (p_2)_k +M_N 
\left(1+ 2 i \theta \alpha_N^{(1)}\theta\gamma_5\right)\right]
\Gamma_\mu(Q^2) \right. \\ \nonumber
 &\times& \left. \left[ E_N(\bp_1)\gamma_0 - i \gamma_k (p_1)_k +M_N
\left(1+ 2 i \theta \alpha_N^{(1)}\theta\gamma_5\right)\right]\right\}_{\alpha\beta}\,.
\label{eq:G3t_SD}
\eea
where $\alpha \beta$ are the Dirac indices.
If we expand in powers of $\theta$ the r.h.s. of eq.~\eqref{eq:GNJNt} we obtain
\be
G_{N J_\mu N}^\theta(\bp_1,\bp_2,x_0,y_0) = G_{N J_\mu N}(\bp_1,\bp_2,x_0,y_0) + i \theta G_{N J_\mu N}^\mcQ(\bp_1,\bp_2,x_0,y_0)\,,
\ee
where
\be
G_{N J_\mu N}(\bp_1,\bp_2,x_0,y_0) = a^6 \sum_{\bx,\by} \e^{i\bp_1\by} \e^{i\bp_2(\bx-\by)} \left\langle \mcN(\bx,x_0) J_\mu(\by,y_0) 
\overline{\mcN}(0)\right\rangle\,,
\label{eq:GNJN}
\ee
is the 3-point function in the standard QCD background, and
\be
G_{N J_\mu N}^\mcQ(\bp_1,\bp_2,x_0,y_0) = a^6 \sum_{\bx,\by} \e^{i\bp_1\by} \e^{i\bp_2(\bx-\by)} \left\langle \mcN(\bx,x_0) J_\mu(\by,y_0) 
\overline{\mcN}(0)\mcQ\right\rangle\,,
\label{eq:GNJNQ}
\ee
contains the insertion of the topological charge evaluated at non-vanishing flow-time $\sqrt{8t}=0.8 r_0$.

Depending on the form factor we are interested in, we can select the appropriate 
Dirac indices with appropriate projectors that we indicate generically as $\Pi$ obtaining
\be
G_\mu^\theta(\bp_1,\bp_2;x_0,y_0;\Pi) = \Tr\left[\Pi G_{N J_\mu N}^\theta(\bp_1,\bp_2;x_0,y_0)\right]\,,
\ee
i.e.
\bea
&\phantom{+}&G_\mu^\theta(\bp_1,\bp_2;x_0,y_0;\Pi) = \frac{{\rm e}^{-E_N(\bp_1)y_0}}{2 E_N(\bp_1)} 
\frac{{\rm e}^{-E_N(\bp_2)(x_0-y_0)}}{2 E_N(\bp_2)} \mcZ_N^*(\bp_1) \mcZ_N(\bp_2) \\ \nonumber
&\times& \Tr\left\{\Pi\left[E_N(\bp_2)\gamma_0 - i \gamma_k (p_2)_k +M_N\left(1+ 2 i \theta \alpha_N^{(1)}\gamma_5\right)\right]
\Gamma_\mu(Q^2) \right. \\ \nonumber 
&\times& \left. \left[ E_N(\bp_1)\gamma_0 - i \gamma_k (p_1)_k +M_N \left(1+ 2 i \theta \alpha_N^{(1)}\gamma_5\right)\right] \right\}\,.
\eea
The spectral decomposition of the correlation functions in eqs.~\eqref{eq:GNJN} and \eqref{eq:GNJNQ} 
traced with a generic projector $\Pi$ are easily obtained
\bea
&\phantom{+}&G_\mu(\bp_1,\bp_2;x_0,y_0;\Pi) = \frac{{\rm e}^{-E_N(\bp_1)y_0}}{2 E_N(\bp_1)} 
\frac{{\rm e}^{-E_N(\bp_2)(x_0-y_0)}}{2 E_N(\bp_2)} \mcZ_N^*(\bp_1) \mcZ_N(\bp_2) \\ \nonumber
&\times& \Tr\left\{\Pi\left[E_N(\bp_2)\gamma_0 - i \gamma_k (p_2)_k +M_N\right]
\Gamma_\mu^{\rm even}(Q^2) \left[ E_N(\bp_1)\gamma_0 - i \gamma_k (p_1)_k +M_N \right] \right\}\,,
\eea
\bea
&\phantom{+}&G_\mu^\mcQ(\bp_1,\bp_2;x_0,y_0;\Pi) = \frac{{\rm e}^{-E_N(\bp_1)y_0}}{2 E_N(\bp_1)} 
\frac{{\rm e}^{-E_N(\bp_2)(x_0-y_0)}}{2 E_N(\bp_2)} \mcZ_N^*(\bp_1) \mcZ_N(\bp_2) \\ \nonumber
&\times& \left\{ \Tr\left[\Pi\left(2 M_N  \alpha_N^{(1)} \gamma_5\right)
\Gamma_\mu^{\rm even}(Q^2) \left( E_N(\bp_1)\gamma_0 - i \gamma_k (p_1)_k +M_N \right)\right] \right. \\ \nonumber
&+& \left. \Tr\left[\Pi\left(E_N(\bp_2)\gamma_0 - i \gamma_k (p_2)_k +M_N \right)
\Gamma_\mu^{\rm even}(Q^2) \left( 2 M_N \alpha_N^{(1)}\gamma_5\right) \right]\right. \\ \nonumber
&+& \left. \Tr\left[\Pi\left(E_N(\bp_2)\gamma_0 - i \gamma_k (p_2)_k +M_N\right)
\Gamma_\mu^{\rm odd}(Q^2) \left( E_N(\bp_1)\gamma_0 - i \gamma_k (p_1)_k +M_N \right)\right] \right\}\,,
\eea
where $\Gamma_\mu^{\rm even}$ and $\Gamma_\mu^{\rm odd}$ are defined in eqs.~\eqref{eq:gamma_evenQ2} and
\eqref{eq:gamma_oddQ2}.
From this expression we already see that the 3-point function with the insertion of the 
topological charge is not directly proportional to the CP-odd form factor $F_3$ but it 
contains additional contributions proportional to $\alpha_N^{(1)}$ and the CP-even form factors.

To extract the form factors traditionally one defines the following chain of ratios
\be
R_\mu^\theta(\bp_1,\bp_2;x_0,y_0;\Pi) = \frac{G_\mu^\theta(\bp_1,\bp_2;x_0,y_0;\Pi)}{C(\bp_2,x_0)}\cdot K(\bp_1,\bp_2;x_0,y_0)\,,
\label{eq:R}
\ee
where $C(\bp,x_0)$, the nucleon 2-point function, is defined as
\be
C(\bp,x_0)={\rm tr}\left[P_+G_{NN}(\bp,x_0)\right] = |Z_N(\bp)|^2 \frac{{\rm e}^{-E_N(\bp) x_0}}{E_N(\bp)}
\left(E_N(\bp)+M_N\right)+ \cdots
\label{eq:C_p}
\ee
and
\be
K(\bp_1,\bp_2;x_0,y_0) = \left[\frac{C(\bp_2,x_0)C(\bp_2,y_0)C(\bp_1,x_0-y_0) }
{C(\bp_1,x_0)C(\bp_1,y_0)C(\bp_2,x_0-y_0)} \right]^{1/2}\,.
\ee

For small $\theta$ we have
\be
R_\mu^\theta(\bp_1,\bp_2;x_0,y_0;\Pi) = R_\mu(\bp_1,\bp_2;x_0,y_0;\Pi) + i \theta R_\mu^\mcQ(\bp_1,\bp_2;x_0,y_0;\Pi)\,,
\ee
where
\be
R_\mu(\bp_1,\bp_2;x_0,y_0;\Pi) = \frac{G_\mu(\bp_1,\bp_2;x_0,y_0;\Pi)}{C(\bp_2,x_0)}\cdot K(\bp_1,\bp_2;x_0,y_0)\,,
\ee
and 
\be
R_\mu^\mcQ(\bp_1,\bp_2;x_0,y_0;\Pi) = \frac{G_\mu^\mcQ(\bp_1,\bp_2;x_0,y_0;\Pi)}{C(\bp_2,x_0)}\cdot K(\bp_1,\bp_2;x_0,y_0)\,.
\ee

Performing the spectral decomposition and retaining only the fundamental state
we obtain 
\bea
&\phantom{+}& R_\mu^\theta(\bp_1,\bp_2;x_0,y_0;\Pi) = \mcN(\bp_1,\bp_2) \\ \nonumber 
&\times& \Tr\left\{\Pi\left[E_N(\bp_2)\gamma_0 - i \gamma_k (p_2)_k +M_N 
\left(1+ 2 i \theta \alpha_N^{(1)}\theta\gamma_5\right)\right]
\Gamma_\mu(Q^2) \right. \\ \nonumber
&\times& \left. \left[E_N(\bp_1)\gamma_0 - i \gamma_k (p_1)_k +M_N 
\left(1+ 2 i \theta \alpha_N^{(1)}\theta\gamma_5\right)\right] \right\}\,,
\eea
where the normalization is given by
\be
\mcN(\bp_1,\bp_2) = \frac{1}{4E_N(\bp_1)E_N(\bp_2)}\left[\frac{E_N(\bp_1)E_N(\bp_2)}{(E_N(\bp_1)+M)(E_N(\bp_2)+M)} \right]^{1/2}\,.
\ee
The ratio~\eqref{eq:R} is defined to remove the leading exponential contributions
and have a plateau for $0 \ll y_0 \ll x_0$ proportional to the form factors.

We can now specialize the projector $\Pi$, the external kinematics, and the current component in order
to compute the form factors we want. For the 2 CP-even form factors we choose
\begin{itemize}
\item $\Pi=P_+$, $\mu=0$, $\bp_1=\bp$, $\bp_2=0$.\\
In this case we indicate $E(\bp)=E$ and $E(\bp_2)=M$
\be
R_0(\bp,\bzero;x_0,y_0;P_+) = \mcN(\bp,\bzero)\cdot 4M_N(E_N(\bp)+M_N)\left[ F_1(Q^2) - \frac{Q^2}{4M_N^2}F_2(Q^2)\right]\,,
\label{eq:R0}
\ee
where
\be
\mcN(\bp,\bzero) = \frac{1}{4M_N}\left[\frac{1}{2E_N(\bp)(E_N(\bp)+M_N)}\right]^{1/2}\,.
\label{eq:npzero}
\ee
Putting everything together, we obtain
\be
R_0(\bp,\bzero;x_0,y_0;P_+) = \left[\frac{E_N(\bp)+M_N}{2E_N(\bp)}\right]^{1/2} 
\left[ F_1(Q^2) -\frac{Q^2}{4M_N^2}F_2(Q^2)\right]\,.
\label{eq:R0b}
\ee
To obtain eq.~\eqref{eq:R0b} and some of the eqs. below we have used the following kinematic relations
\be
\left|\bq\right|^2=\left|\bp\right|^2 = E_N^2 - M_N^2\,,
\ee
and
\be
q^2 = (M_N-E_N)^2-\left|\bq\right|^2 \Rightarrow q^2=(M_N-E_N)^2 - (E_N^2 -M_N^2)  \Rightarrow q^2= 2M_N(M_N-E_N) <0\,,
\label{eq:q2def}
\ee
This implies that 
\be
E_N-M_N = -\frac{q^2}{2M_N} = \frac{Q^2}{2M_N}\,.
\ee
Any of the relations in eq.~\eqref{eq:q2def} defines the $Q^2=-q^2$ to be used when analyzing the form factors.

\item $\Pi=i P_+ \gamma_5 \gamma_j$, $\mu=i$, $\bp_1=\bp$, $\bp_2=0$.\\
After some algebra we obtain
\be
R_i(\bp,\bzero;x_0,y_0;i P_+ \gamma_5 \gamma_j) = \mcN(\bp,\bzero)\cdot 4 M q_k \epsilon_{ijk} \left[ F_1(Q^2) + F_2(Q^2)\right]\,,
\ee
and using the expression for the normalization~\eqref{eq:npzero}, we obtain
\be
R_i(\bp,\bzero;x_0,y_0;i P_+ \gamma_5 \gamma_j) = \left[\frac{1}{2E_N(\bp)(E_N(\bp)+M_N)}\right]^{1/2} q_k \epsilon_{ijk} \left[ F_1(Q^2) + F_2(Q^2)\right]\,,
\label{eq:Ri}
\ee

\end{itemize}

For the CP-odd form factor there are several choices for the Dirac projector.
The analysis presented in this paper uses
\begin{itemize}
\item $\Pi=i P_+ \gamma_5 \gamma_i $, $\mu=0$, $\bp_1=\bp$, $\bp_2=0$.\\
If we compute the coefficient of $i \theta$, we obtain after some algebra
\bea
\label{eq:R05}
R_0^\mcQ(\bp,\bzero;x_0,y_0;i P_+ \gamma_5 \gamma_i) &=& \mcN(\bp,\bzero)\cdot \\ \nonumber
&\phantom{+}&\left\{ -4 \alpha_N^{(1)} M_N q_i \left[ F_1(Q^2) + \frac{E_N+3M_N}{2M_N}F_2(Q^2)\right] \right.\\ \nonumber
&-& \left.2(E_N + M_N) q_i F_3(Q^2)\right\}
\eea
\end{itemize}
Here the importance of a precise determination of $\alpha_N^{(1)}$ becomes clear. The mixing of parity states induces
spurious contributions to the correlation functions proportional to the CP-even form factors. 
These contributions need to be subtracted in order to determine the nucleon EDM.

\end{appendix}

\bibliography{edm_theta}      
\bibliographystyle{h-physrev}
\end{document}